\documentstyle[mymulticol,pra,aps,amsfonts]{revtex}

\newcommand{\beq}{\begin{equation}}
\newcommand{\eeq}{\end{equation}}
\newcommand{\beqy}{\begin{eqnarray}}
\newcommand{\eeqy}{\end{eqnarray}}

\newtheorem{Definition}{Definition}
\newtheorem{Lemma}{Lemma}
\newtheorem{Theorem}{Theorem}
\newtheorem{Corollary}{Corollary}

\newenvironment{Proof}{{\it Proof: \,}}{$\Box$ \vspace{0.3cm}}

\newenvironment{Definition*}{{\bf Definition}}{}

\makeatletter
\def\@beginTheorem#1#2{\trivlist \item[\hskip \labelsep{\bf #1\ #2}]}
\def\@opargbegintheorem#1#2#3{ \trivlist
      \item[\hskip \labelsep{\bf #1\ #2\ (#3)}]}
\makeatother

\makeatletter
\def\@beginLemma#1#2{\trivlist \item[\hskip \labelsep{\bf #1\ #2}]}
\def\@opargbeginLemma#1#2#3{ \trivlist
      \item[\hskip \labelsep{\bf #1\ #2\ (#3)}]}
\makeatother

\makeatletter
\def\@beginDefinition#1#2{\trivlist \item[\hskip \labelsep{\bf #1\ #2}]}
\def\@opargbeginDefinition#1#2#3{ \trivlist
      \item[\hskip \labelsep{\bf #1\ #2\ (#3)}]}
\makeatother

\makeatletter
\def\@beginCorollary#1#2{\trivlist \item[\hskip \labelsep{\bf #1\ #2}]}
\def\@opargbeginCorollary#1#2#3{ \trivlist
      \item[\hskip \labelsep{\bf #1\ #2\ (#3)}]}
\makeatother

\makeatletter
\def\@beginExample#1#2{\trivlist \item[\hskip \labelsep{\bf #1\ #2}]}
\def\@opargbeginExample#1#2#3{ \trivlist
      \item[\hskip \labelsep{\bf #1\ #2\ (#3)}]}
\makeatother

\def\C{{\mathbb{C}}}

\def\R{{\mathbb{R}}}

\def\N{{\mathbb{N}}}

\newcommand{\cH}{{\cal H}}

\newcommand{\cS}{{\cal S}}
\newcommand{\cT}{{\cal T}}
\newcommand{\tl}{\tilde}
\newcommand{\T}{{\cal T}}
\newcommand{\E}{{\cal E}}

\title{The thermodynamic cost of reliability and low temperatures:\\
Tightening Landauer's principle and the Second Law}

\author{D. Janzing\thanks{Electronic address: janzing@ira.uka.de}, P. Wocjan, R. Zeier, R. Geiss, and Th. Beth}
\address{Institut f\"ur Algorithmen und Kognitive Systeme, Am Fasanengarten 5,
    D--76\,128 Karlsruhe, Germany}

\begin{document}
\maketitle

\begin{abstract}
Landauer's principle states that the erasure of one bit of information
 requires the 
free energy $kT\ln 2$. We argue that the reliability of the bit erasure
process is bounded by the accuracy inherent in the statistical state
of the energy source (`the resources')
driving the process. We develop a general framework describing
the `thermodynamic worth' of the resources with respect to 
{\em reliable bit erasure} or {\em good cooling}. This worth turns out to be given
by the distinguishability of the resource's state from its equilibrium state
in the sense of a statistical inference problem.
Accordingly, Kullback-Leibler relative information is 
 a decisive quantity for the `worth' of the resource's
state. 
Due to the asymmetry of relative information,
the reliability of the erasure process
is  bounded rather by 
the relative information
of the {\em equilibrium state with respect to  the actual state}
 than   by
the relative information of the {\em actual state with respect to
the equilibrium state} (which is the free energy up to constants).
\end{abstract}

\begin{multicols}{2}

\section{Introduction}

One of the characteristic features of technological progress 
is the increase of human ability to control and design
the microscopic world. Especially the recent successes in manipulating
simple quantum systems (for example in the  context of Quantum Computing
research) are one aspect of this general development.
Since every process  controlling microscopic particles
is disturbed by heat, this progress is strongly connected with
the invention of efficient {\em cooling mechanisms} (see \cite{AAKVC},
\cite{MECZ},\cite{MCLZ}). 
This statement is in some sense\footnote{We use the cautious formulation `in some sense'
because  of the 
 following objection: If the system has a large energy gap between 
its ground state and the first excited  state, it is
in an almost pure state even for not too low temperatures.} a {\em tautological} one:
Preparing a physical system in a {\em pure} quantum
 state means preparing a state {\em without entropy}, i.e.,
a system {\em without heat}. 
In present day cooling techniques, the size of the required apparatus  
is quite impressive compared to the tininess of the cooled
systems. In contrast, miniaturization in computer technology 
will require smaller, efficient and power saving mechanisms for draining off
entropy on the nanoscopic or microscopic level. 
This raises the question for fundamental lower bounds on the resources
needed for cooling simple quantum systems.
At first sight
the answer seems to be  given by well-known
  thermodynamic theory, in particular
 the Second Law: Extracting the entropy $S$ from a system
requires the energy $S\, k T$ where $k$ is  Boltzmann's constant
and $T$  the temperature of the surrounding  heat bath absorbing the
 entropy.
Another formulation of this law is Landauer's principle saying that
the erasure or initialization of one bit being in an  unknown state
requires at least the energy $\ln 2 \, k T$ (see \cite{Be}, \cite{La1},
 \cite{La2}).
But this cannot be the complete answer:  To understand
the fundamental limitations on scaling down the cooling apparatus
and reducing the resources, we 
model the cooling process as an energy conserving 
 unitary dynamics on the composition
of the considered quantum system with another one (`the resources').
Within this microscopic model we do not expect that
necessary {\em and sufficient} conditions for the
  resource's quantum  state
to enable effective cooling procedures 
are given by  well-known laws of thermodynamics. 

Of course, a lot of steps have already been made towards a refinement
of thermodynamics on the level of low-dimensional quantum systems 
(see e.g. \cite{AC}, \cite{Pl}). Actually,
 one should 
reckon all the
 results concerning  entanglement purification \cite{BBPS}, quantum  error correction \cite{Ve}\cite{BDSW}, quantum data compression \cite{Ho},
and logical cooling \cite{SV}
as such since they are dealing essentially with the {\em transport}
and {\em concentration} 
 of information by operations on compositions
of simple quantum systems.
Nevertheless, our approach is rather different from those ones:
Our microphysical models of cooling
 include the energy source -- a quantum system as well --
driving the process, i.e., we restrict
the class of unitary transformations to those conserving the total
Hamiltonian of the system. This setup emphasizes the fact, that we
want to
develop a theory of {\em thermodynamics} in contrast to a pure 
theory of {\it information}: 
The latter one deals with {\em information} only, while the first one
focuses  on  the 
{\em relation between
energy and information}.

Some consequences of the restriction to energy conserving transformations
can be illustrated easily: 
Consider a bipartite quantum system consisting of a harmonic oscillator
 with frequency $\nu$ and a two-level system with energy levels $0$ and $h\nu$.
Assume both systems to be in their equilibrium states for the same 
(finite) temperature.
Then one can easily construct unitary transformations on the composite
 Hilbert space extracting entropy from the two-level system
and pumping into the oscillator. One can even show, that there are
no bounds on the efficiency of such a cooling process, i.e., the
state of the two-level system can be prepared arbitrarily close
to a pure one.
In contrast, there is no {\em energy conserving} unitary transformation
changing the state of the system at all. Such a process would
even violate the Second Law, since this would be a dynamics
 producing free energy without the use of an additional energy source.
Accordingly, if the state of the harmonic oscillator differs {\em slightly}
from its equilibrium state we will expect that an energy preserving
process can only have a {\em slight} cooling effect.
Lead by this intuition, we investigate in which way the size of the
deviation of the quantum system's state from its equilibrium state
 determines its
`thermodynamic worth' for enabling good cooling processes, or more
generally, for {\em precise} preparation of quantum states.
Reformulated
in the spirit of the `thermodynamics of computation', we investigate
the minimal resource requirements for a {\em reliable} bit erasure process.

The paper is organized as follows: 
In section II we give a short introduction into thermal equilibrium states
of quantum systems. In section III we present the formal setup
of the microscopic cooling process  and give necessary
and sufficient conditions for the resource's state to allow for cooling
 a two-level system.
In section IV we introduce a more flexible model 
in which cooling is described by a unitary dynamics on a tripartite system:
The resources, the environment being in thermal equilibrium, and
the two-level system to be cooled down.
 We prove that cooling is possible if and
only if the time average of the resource's state does not agree
with its equilibrium state. 
If the temperature of the two-level system is already below 
the environment's temperature, the deviation of the resource's state
from equilibrium determines whether it is possible to cool the qubit
even more.
The second part of this section answers the
totally different question of the lowest qubit-temperature
 which can be obtained by using the given resources if the qubit
has initially the same temperature as the environment.
We show that the determination of the lowest obtainable temperature
can be reduced  to a quantum inference problem, namely the determination
of error probabilities of a decision rule for distinguishing the 
resource's state from its equilibrium state.
Sections V-VII shows consequences of our theory and analyze
in which sense they go beyond well-known laws of thermodynamics.

\section{Thermodynamic background}

Let $\cH$ be the finite or infinite dimensional  Hilbert space of 
a quantum system and $H$ a selfadjoint operator acting on $\cH$ representing
its Hamiltonian. Then, for any temperature $T$ the density matrix 
\[
\rho_T:=e^{-H/(kT)}/tr(e^{-H/(kT)}),
\]
where $k$ is Boltzmann's constant,
is called the {\em thermal equilibrium state} with temperature $T$
 provided that $tr(e^{-H/(kT)})$
exists. Note that we 
do not define temperature as a property of every state, but 
 merely for those of the form described above.

As usual, we will use the inverse temperature defined
by $\beta:=1/(kT)$ and consider the class of states
\[
\rho_\beta:=e^{-\beta H}/tr(e^{-\beta H})
\]
for any $\beta$ with $-\infty \leq \beta \leq  \infty$.

In the special  case of a non-degenerate two-level system this implies that
an inverse temperature can be assigned to any 
density matrix commuting with the Hamiltonian. 
For two diagonal-states the state with lower
$\beta$ is the hotter state. The fact, that  {\em heating} up to a value
$\beta <0$ {\em decreases} the entropy is the well-known phenomenon
of temperature inversion \cite{Go}. In order to avoid confusion we emphasize that 
{\em heating} 
 means here  {\em increasing} the  occupation probability for the
 upper
 state. This is connected with an {\em increase} of entropy for $\beta>0$ and a
{\em decrease} of entropy for $\beta<0$.
This unusual connection between entropy and heat due to temperature inversion
might be confusing. However, we will mostly focus on
cooling, since the corresponding statements
for heating in our sense can be obtained analogously.
In contrast, if one considers the maximally
mixed state ($T=\infty$) as the hottest one, there is no 
such analogy and the preparation of the hottest states does not cause
any difficulties comparable to the preparation of the coldest one.  

Since we want to interpret our results in the context of
`thermodynamics of computation' we keep in mind that 
 a two-level system can be considered  as an
one-bit-memory and  any process producing an (almost) pure state
from a mixed one will be considered as   
an  {\em erasure} process of one unknown bit of information. 

In  the following sections the dependence of the equilibrium states
from the temperature will  mostly  not be mentioned explicitly, since 
troughout the paper we fix
one common reference temperature $T\neq 0, T\neq \infty$
(and the corresponding inverse temperature $\beta$)
 representing the temperature of
the particle's environment.

\section{The model}

To investigate the ability of cooling or heating
 a multi-level quantum system within a precise mathematical framework,
 we introduce 
some terminology:
Here, a {\em quantum system} is uniquely characterized by
its Hamiltonian $H$, since it determines in a unique way
 the corresponding Hilbert space
and its dynamics. Up to an irrelevant translation of the energy scale,
for any fixed inverse temperature $\beta$ there is an 
one-to-one correspondence between the system's Hamiltonian and
its equilibrium state. Note that any unitary operator
$u$ commutes with $H$ if and only if it commutes with 
its equilibrium state provided that $\beta\neq 0 $ and $\beta\neq \infty$, i.e,
a dynamics is {\em energy conserving} if and only if it
preserves the equilibrium state.
 
Every quantum system can be in different statistical 
states, described by a density matrix
acting on the same Hilbert space as the Hamiltonian.
We call a system being in a particular statistical state an {\em object}.
More formally we define:

\begin{Definition}\label{object}
~\newline
\begin{enumerate}
\item
A (quantum) {\bf system} is described by a density matrix $\gamma$
(its `equilibrium state') acting on a finite dimensional Hilbert space $\cH$.

\item An {\bf object} is a pair $(\rho,\gamma)$ where $\rho$ is a
density matrix describing the actual mixed state of the system.
Hence, a system in equilibrium is described by an object of the form
\[
(\gamma, \gamma).
\]
Usually we will assume both matrices to have full rank.

\item
For two systems $\gamma$ and $\tilde{\gamma}$ we define the
 {\bf composed system}
as the system  determined  by the equilibrium state
\[
\gamma\otimes \tl{\gamma}.
\]

\item
For two objects $O:=(\rho,\gamma)$ and $\tilde{O}:=(\tilde{\rho},\tilde{\gamma})$
the {\bf composed object} is defined to be 
\[
O\times \tilde{O}:=
(\rho\otimes \tilde{\rho},\gamma\otimes \tilde{\gamma})
\]

\item
If $u$ is a unitary operator acting on $\cH$ with $u\gamma u^*=\gamma$,
 i.e, $u$ is an `energy conserving
reversible dynamics', we define
the {\bf allowed transformation} $T_u$ on the object $O:=(\rho,\gamma)$
 by:
\[
T_u((\rho,\gamma)):=(u\rho u^*,\gamma).
\]
In abuse of language,
 we will call $u$ an allowed transformation as well.

\item
If a system $\gamma \otimes \tilde{\gamma}$ is in the state
$\rho$  (where $\rho$ is not a tensor product state necessarily),
we define the {\bf restriction} of the object $O:=(\rho,\gamma \otimes
\tilde{\gamma})$ to its left, respectively right, component as
\[
O_l:=(tr_r(\rho),\tilde{\gamma})
\]
and
\[
O_r:=(tr_l(\rho),\gamma ),
\] 
where $tr_l$ and $tr_r$ denote the partial trace over the left, repectively
right, component in the tensor product.
\end{enumerate}
\end{Definition} 

Within this framework, the problem of cooling
a two-level system (`qubit') by given resources can be formalized
as follows:

Given the arbitrary object $O$ (`the resources') and the qubit
$Q:=(\sigma,\sigma)$, with
\[
\sigma:= \frac{1}{1+e^{-\beta E}} \, diag (1,e^{-\beta E}),
\]
where $E$ is the energy gap of the two-level system.
Find an allowed transformation $T_u$ on
\[
O\times Q
\]
which serves as a cooling process for $Q$, i.e., 
\[
(T(O\times Q))_r
\]
is a qubit with a lower or higher temperature compared to the initial
state $\sigma$.

Firstly we will look for those allowed transformations  which
minimize or maximize the occupation probability for the
upper level. 
Let $\sigma_z$ be the Pauli matrix
\[
\sigma_z:= \,diag (1,-1)
\]
and assume the Hamiltonian of the qubit
to be
\[
\tilde{H}:=\, diag(0,E).
\]
Then the occupation probability for the
upper level is maximized (respectively minimized) for
those transformations $u$ which minimize (respectively
maximize) the term   
\[
tr( u (\rho\otimes \sigma) u^* (1\otimes \sigma_z)).
\]

We find necessary and sufficient conditions for the transformations
 $u$ to be optimal:

\begin{Lemma}\label{Bed}
Let $\alpha$ be the density matrix of a bipartite system
composed of a qubit with equilibrium state $\sigma$ as above
and another arbitrary system with equilibrium state $\gamma$, i.e.,
we have the object 
\[
(\alpha, \gamma\otimes \sigma).
\]
Let $P_i$ be the spectral projections of $\gamma\otimes \sigma$.
Assume that the following two conditions hold:

\begin{enumerate}
\item All the operators $P_j\alpha P_j$ commute with $1\otimes \sigma_z$,
\vspace{0.3cm}

     so that we can divide the eigenvalues of the 
    restriction of $P_j\alpha P_j$ to the range of $P_j$ into
    subsets $\Gamma^j_+$ and $\Gamma^j_-$  corresponding to the
    eigenvectors 
    of $1\otimes \sigma_z$ with eigenvalues $+1$ and $-1$,
    respectively.
 
\item The smallest eigenvalue in $\Gamma^j_+$ is greater than the greatest
        eigenvalue in $\Gamma_-^j$.

\end{enumerate}

Then there is no allowed transformation $u$
on $O\times Q$ decreasing the occupation probability
of the upper state, i.e., we have:
\[
tr (u\alpha u^* (1\otimes \sigma_z)) \geq tr (\alpha (1\otimes \sigma_z)),
\]
for every unitary operator $u$ with $[u,\gamma\otimes \sigma]=0$.
\end{Lemma}

The Lemma will be proved in the appendix.

For any allowed transformation $u$ we can decide whether there
can exist a better one for cooling by setting
$\alpha:=u (\rho\otimes \sigma)u^*$. Then Lemma \ref{Bed}
gives a criterion whether there can exist a better transformation $u'$.
Furthermore it shows, that the optimal transformation 
for reducing the probability of the upper state or the lower state
can always be chosen in such a way that the reduced  density matrix
of the qubit
is still  diagonal after one has performed the unitary transformation $u$.
Therefore we can obtain an equilibrium state with a temperature
different from the reference temperature.

We shall use the following notion:

\begin{Definition}\label{tauglichDef}
Let $Q:=(\sigma,\sigma)$ be a qubit in its equilibrium state.
We say `the object $O:=(\rho,\gamma)$' can be used for cooling
$Q$ if there is an allowed transformation $T_u$ on
$O\times Q$ such that
\[
tr (u (\rho\otimes \sigma)u^* (1\otimes \sigma_z))> tr( \rho \sigma_z).
\]
We say that it can be used for heating if we have `$<$' instead of `$>$'.
\end{Definition}

In order to give necessary and sufficient conditions for the 
possibility of cooling or heating the following suggestive definition
turns out to be useful:

\begin{Definition}\label{relativTemp}
For any object $(\rho,\gamma)$ let
 $|i\rangle$ and $|j\rangle$ be eigenvectors of $\gamma$.
Let $E_i$ and $E_j$ be the corresponding eigenvalues of the system's
 Hamiltonian, i.e., 
\[
E_j-E_i= \frac{(\ln \langle i|\gamma |i\rangle -\ln \langle j|\gamma |j\rangle)
}{\beta}. 
\]
Then the {\bf relative inverse temperature} with respect to the states
$|i\rangle$ and $|j\rangle$ is defined to be
\[
\beta_{|i\rangle, |j\rangle}:= \frac{\ln \langle i|\rho |i\rangle -
\ln \langle j|\rho |j\rangle}{E_j-E_i}.
\]
Similarly, we define the relative temperature
\[
T_{|i\rangle,|j\rangle}:=\frac{1}{k \beta_{|i\rangle,|j\rangle}}.
\]
\end{Definition}

Using this definition  we have an easy criterion for the possibility
of cooling:

\begin{Theorem}\label{Theorelativ}
An object $O:=(\rho,\gamma)$ can be used for cooling a qubit 
$Q:=(\sigma,\sigma)$ with energy gap $E$ and 
the inverse temperature $\beta$ if and only if there
is a pair $|i\rangle$ and $|j\rangle$ of eigenvectors of
the Hamiltonian $H$ (corresponding to $\gamma$)
with different eigenvalues  $E_i$ and $E_j$
such that $E_i-E_j=E$ and
\[
\beta_{|i\rangle,|j\rangle}> \beta.
\]
\end{Theorem}

\begin{Proof}
Assume $\beta_{|i\rangle,|j\rangle}>\beta$.
Let $|1\rangle,\dots ,|l\rangle$ be a basis of eigenvectors
of $\gamma$.
Let $|0\rangle$ and $1\rangle$ be the lower and upper state
of the two-level system (In case of degenerated levels the choice is
irrelevant).  
 Then the occupation probability for the
ground state is given by
\[
\langle 0|\sigma |0\rangle
=\sum_j \langle j|\rho|j\rangle \langle 0|\sigma |0\rangle.
\]
Now we perform the transformation $u$ by permuting the states
by the involution
\[
|i\rangle\otimes |0\rangle \,\,\,\leftrightarrow \,\,\,
 |j\rangle \otimes |1\rangle
\]
and acting trivial on the other tensor product basis states.
 The probability for the lower state is changed by the amount
\begin{eqnarray*}
&&tr ((u(\rho\otimes \sigma) u^* -(\rho\otimes\sigma))( 1\otimes |0\rangle\langle 0|))\\&=&
\langle j|\rho|j\rangle \langle 1|\sigma |1\rangle
-\langle i|\rho|i\rangle \langle 0|\sigma |0\rangle
.
\end{eqnarray*}
The latter term is negative 
by assumption and
due to the definition of $\beta$ and $\beta_{|i\rangle,|j\rangle}$.

Assume $\beta_{|i\rangle, |j\rangle}\leq \beta$.
Clearly, for any $j$ the spectral projection $P_j$
can be written as
\[
P_j= (Q_+\otimes |0\rangle \langle 0|) \oplus 
(Q_-\otimes|1\rangle \langle 1|),
\]  
where $Q_+$ and $Q_-$ are spectral projections of $\gamma$.
Since $\sigma$ commutes with $|0\rangle\langle 0|$ and 
$|1\rangle\langle 1|$ we
have:
\begin{eqnarray*}
P_j (\rho\otimes \sigma) P_j&=&Q_+ \rho Q_+ \otimes |0\rangle\langle 0|
 \langle 0|\sigma | 0\rangle \oplus \\
&&Q_-\rho Q_-\otimes |1\rangle\langle 1| \langle 1|\sigma |1\rangle.
\end{eqnarray*}
The eigenvalues of the first component in this direct sum belong
to $\Gamma^j_+$, those in the second to $\Gamma^j_-$.
If $E_i-E_j=E$ the quotient of any eigenvalues of
$Q_+\rho Q_+$ and any eigenvalue of $Q_-\rho Q_-$ can never exceed
$e^{-\beta_{|i\rangle,|j\rangle} E}$. Therefore $\beta_{|i\rangle, |j\rangle}\leq
\beta$ implies that condition (2) in Lemma \ref{Bed} is fulfilled.    
\end{Proof}

In the sense of  the definition $\ref{relativTemp}$ we have the
 strong statement, that
the low temperature which {\em should be attained} in the qubit must 
already be {\em inherent in the used resources}.
For the moment, the problem of cooling seems to be circular and
one might ask, why cooling  is possible at all.

We will show that there is an easy answer, since
arbitrary low relative temperatures can be obtained by
composing {\em  many} objects deviating from their equilibrium state. 
In particular, the composition of two objects $O_1$ and $O_2$ 
 being  in their thermal equilibrium states for the inverse 
temperature $\beta_1$ and $\beta_2$, respectively can contain
inverse temperatures larger than $\beta_1$ and $\beta_2$.
This is the quantum analogue of the well-known fact from
classical thermodynamics, that {\em cooling} can be driven by {\em heat}
without any other energy supply.
This principle is used in an absorption heat pump for instance. 

This indicates that the calculation of the
 relative temperatures obtained by {\em composing}  objects
might give interesting insights in the problem of 
`the origin of low temperatures'. 
We will develop a quite general theory of relative temperatures
in composed systems, but
we will restrict our investigations
to the case that the density matrices of the considered systems
are diagonal with repect to any basis diagonalizing the Hamiltonian.
Furthermore we will restrict the class of allowed transformations
to those which  permute the basis states.
We will call this the `quasi-classical case' 
and define:

\begin{Definition}
~\newline
\begin{enumerate}
\item
A {\bf quasi-classical (l-level) system} is described by 
a vector $g\in \R^l$
defining the probabilities for finding the system in one of the
states $\{1,\dots,l\}$.

\item
A {\bf quasi-classical object} is a pair $(p,g)$ where 
$p\in \R^l$ is the probability distribution of the actual state
and $g\in \R^l$ the equilibrium distribution. 
Let $p_i$ and $g_i$ be the components of the vectors $p,g$.

\item
An {\bf allowed transformation} is a permutation $\pi$ of the states
$1,\dots,l$ which leaves $g$ invariant, i.e.,
$g_{\pi(i)}=g_i$ for every $1\leq i\leq l$.

\item
Composition of objects and composition and restriction of systems
are defined  as in the quantum case (see Definition \ref{object}), i.e.,
we have tensor product vectors describing joint probability
distributions, restrictions of objects are defined by marginal distributions.

In analogy to Definition \ref{relativTemp}, a relative inverse temperature
 $\beta_{i,j}$ can be assigned to  any pair $(i,j) \in \{1,\dots,l\}^2$.
\end{enumerate}
\end{Definition}

Now we are able to give an example for the statement that
 the composition of
an $n$-fold copy of the identical object
can lead to
arbitrary low temperatures as $n$ increases:
Take a system with the energy levels $0,E,2E$ being in the statistical state
$p=(p_1,p_2,p_3)$.
Let $n$ be an odd number and set $n=2l-1$. 
We assume
\[
1> \frac{p_3}{p_2}\frac{p_1}{p_2}=:d
\]
In the $n$-fold composition of the object $(p, g)$, i.e., in
$(p^{\otimes n},g^{\otimes n})$, we
consider
the following two states $|1\rangle$ and $|2\rangle$:

Let $|1\rangle$ be  
 some state  in which $l$ of the subsystems
are on the level $2E$ and $l-1$ are in the level 0. Let $|2\rangle$ be
the unique state  where every system has energy $E$.
The quotient of the probabilities of these two states
is
\[
d^l\frac{p_2}{p_1},
\]
the energy difference of both is $E$.
Hence we get the relative inverse temperature
\[
\beta_{1,2}
=-\frac{1}{E} (l \ln d + \ln (p_2/p_1))\, ,
\]
which tends to infinity for increasing $l$.

It turns out, that the problem of determining the relative inverse temperatures
in an object composed of two quasi-classical ones is a geometrical one:
For any pair $(i,j)$ of states of the object $O:=(p,g)$  we define
a vector $v_{i,j} \in \R^2$ by
\beq\label{vDef}
v_{i,j}(O):=(\frac{1}{\beta}\ln (g_i/g_j),\ln (p_i/p_j)).
\eeq
Note that the quotient of relative inverse temperature
and the reference inverse temperature $\beta$ of the pair $(i,j)$ is given
by the tangens of the angle enclosed by $v_{ij}$ and the x-axis. 
In any composed object $O\times \tl{O}$
we denote the state $(i,\tilde{i})$ by 1 and the state
  $(j,\tilde{j})$ by 2.
We obtain
\[
v_{1,2}(O\times \tl{O})=v_{i,j}(O)+v_{\tl{i},\tilde{j}}(\tl{O})
\]
as the sum of the corresponding vectors for the subsystems.

If we define $V_O:=\{ v_{i,j} (O) |   1\leq i,j \leq l\}$, we
get:
The inverse temperatures available in the $n$-fold of the object
are given by the possible values of  
$\tan \phi$, where $\phi$ is the angle enclosed by the vector
\[
\sum_{i=1}^n x_i
\]
and the x-axis and  $x_i$ are arbitrary vectors taken from the set  $V_O$.

Therefore the problem of finding the lowest relative temperature
in a composed object is a geometrical one.

\section{Including the environment}

The problem of finding the lowest relative temperature in a given object
 is a little bit artificial for two reasons:
Firstly, the optimal pair of states can only be used for
cooling those two-level systems which have  the same energy gap.
Of course  it would be more natural 
to fix the required energy gap in advance.
But then, in the generic case, one will not find any appropriate pair of states at all.
Secondly, it does not make sense to assume that 
the two-level system and the resources must be isolated
from the rest of the world. Since it is even impossible to {\em prevent}
this systems from interacting with the rest of the world, it 
seems unphysical  to {\em forbid}  such an interaction even if it
would {\em help} for cooling.

In a modified model,
both shortcomings of the theory can be removed at once:
We will investigate the possibilities of cooling
 a given object under the assumption that one can use the help of
arbitrary additional equilibrium objects. They can be thought of as 
the system's environment, i.e.,
physical systems as particles and field surrounding
the considered objects. 
We will assume the environment to be in its equilibrium state,
since we consider this as its {\em defining} property: 
every non-equilibrium object would be reckoned as 
additional {\em resources}.
 
In other words, we will investigate the `worth' of the resources
with respect to cooling under the assumption that
{\em equilibrium} objects can be obtained {\em for free}.
There are two reasons why the inclusion of ancilla
equilibrium objects may help for cooling:
On the one hand, generically the energy gaps of the resources pure states
will not coincide   with the energy difference of the two-level system.
Then, an additional equilibrium object with an appropriate level structure
enables to perform nontrivial transformations at all. 
On the other hand, a {\em cooling} process driven by {\em heat}
without any other energy supply
is only  possible with the use of objects having the 
(lower) reference temperature. Loosely speaking:
The Second Law states that
`heat without cold is worthless' for driving any process.

In order to avoid unnecessary  mathematical
complications, we will assume
the ancilla objects to be {\em finite} dimensional quantum systems.
This should not be considered as  an essential  restriction, since we are only
interested in  statements
which do not refer to any particular level structure of the ancillas.
Furthermore, in some sense the infinite dimensional case is included 
in our analysis, since we allow sequences of systems with growing
 dimension
as environments. 

We shall see that the help of an appropriate environment is so useful, that
in the quasi-classical case even {\em every} non-equilibrium object
enables cooling.
For the quantum case we will show, that
an object enables cooling if and only if the {\em time average}  of 
its state differs from the equilibrium state. 
The time average $\overline{\rho}$ is defined by
\[
\overline{\rho}:=\lim_{t\to\infty} \frac{1}{t}\int_0^t e^{-iHs} \rho
 e^{iHs}ds.
\]
It is given by
\[
\overline{\rho}=\sum_j  P_j\rho P_j
\]
where $P_j$ are the spectral projections of the system's Hamiltonian $H$ (and 
the corresponding equilibrium state $\gamma$).
This can be seen by theorems from ergodic theory on Hilbert spaces
\cite{Kr}: By taking the trace as an inner product on the space of matrices,
the time evolution is unitary on the density matrices and the map
$\rho \mapsto \sum P_j \rho P_j$ is the orthogonal projection
on the eigenspace of the generator $i[H,.]$ with eigenvalue $0$.

If an object enables cooling of a qubit having environment's temperature,
 it is natural to ask whether the object allows cooling even if the qubit
is already colder than the environment. 

 We define the {\em lower} (respectively upper)
{\em limit temperature} of a resource object as  the greatest 
(respectively lowest)
initial temperature of the qubit such that cooling (respectively heating)
is just impossible. Note that the reference temperature, i.e., the temperature
of the used ancilla objects, 
is fixed however.
We will introduce a parameter 
which will turn out to
determine the lower and the upper limit temperatures at once:

\begin{Definition}\label{mddDef}
For any object $O:=(\rho,\gamma)$ and any pair $|i\rangle$, $|j\rangle$ of
eigenstates of $\gamma$ with eigenvalues $\lambda_i,\lambda_j$ we set:
\[
f_{|i\rangle,|j\rangle}(O):=  
\ln \langle i|\rho|i\rangle -\ln 
\langle j|\rho|j\rangle -\ln \lambda_i +\ln \lambda_j .
\]
Then we define
the {\em maximal
diagonal deviation} from equilibrium as
\beq\label{mdd}
D(O):=\max\{|f_{|i\rangle, |j\rangle }|\},
\eeq
where the maximum is taken over all pairs of eigenstates.
\end{Definition}

Obviously $D(O)=0$ if and only if 
the diagonal entries 
of $\rho$ agree with the entries of the equilibrium state
with respect to every basis diagonalizing $\gamma$.
 This justifies the terminology.
Easy considerations show, that 
$
D((\rho,\gamma))=0
$
if and only if $\overline{\rho}=\gamma$, where $\overline{\rho}$ is the 
time average of $\rho$.
Furthermore we have the following reformulation of Definition
\ref{mddDef}:

\begin{Lemma}\label{Umformulierungmdd}
Let $O:=(\rho,\gamma)$ be an arbitrary object and
 $P_i$ be the spectral projections of $\gamma$ for the eigenvalues
$\lambda_i$. Then
the maximal diagonal deviation is given by:
\begin{eqnarray*}
&&D(O)=\\&&\max_{i,j}\{|\ln (\|P_i\rho P_i\|)+\ln (\|(P_j\rho P_j)^{-1}\|)
- \ln \lambda_i + \ln \lambda_j|\},
\end{eqnarray*}
where $(.)^{-1}$ denotes the pseudoinverse of any matrix and $\|.\|$ is the 
operator norm defined by $\|a\|:=\max_x\{\|ax\|/\|x\|\}$ where $\|x\|$ is the
euclidean norm of the vector $x$.
\end{Lemma}

\begin{Proof}
Obviously, Definition \ref{mddDef} can be reformulated as 
\begin{eqnarray*}
&&D(O)=\\&&\max_{i,j}\max_{|\psi\rangle,|\phi\rangle}\{ 
 |\ln (\langle \psi|\rho|\psi\rangle) -\ln (\langle \phi|\rho|\phi\rangle) -\ln \lambda_i+ \ln \lambda_j|\},
\end{eqnarray*}
where $|\psi\rangle $ and $|\phi\rangle$ have to be eigenvectors 
of $\gamma$ corresponding to $\lambda_i$ and $\lambda_j$,
 respectively.  
The term in the braces is maximized if $|\psi\rangle$ is the eigenvector
of $P_i\rho P_i$ corresponding to its largest eigenvalue and
$|\phi\rangle$ corresponding to the smallest eigenvalue of $P_j\rho P_j$.
But then one has:
\[
\langle \psi |\rho|\psi\rangle =\|P_i\rho P_i\|
\]
and
\[
\langle \phi |\rho|\phi \rangle=\|(P_j\rho P_j)^{-1}\|^{-1}.
\]
\end{Proof}

The maximal diagonal deviation is a superadditive quantity, for 
 quasi-classical objects it is only additive:

\begin{Theorem}
We have
\[
D(O\times \tilde{O})\geq D(O)+D(\tilde{O})
\]
for  arbitrary objects  $O:=(\rho,\gamma)$ and 
$\tl{O}:=(\tl{\rho},\tl{\gamma})$ with equality if\footnote{There are easy examples, showing that this
condition cannot be dropped:
Take a qubit with $diag (0,E)$ as Hamiltonian and a coherent superposition
of $|0\rangle$ and $|1\rangle$ such that the diagonal entries
of the corresponding density matrix agree with the equilibrium distribution.
Hence $D$ vanishes for this object. But the composition of two such objects
has non-vanishing $D$.}
 $\rho$  commutes with $\gamma$ {\em or} $\tilde{\rho}$ commutes with
  $\tilde{\gamma}$.
\end{Theorem}

\begin{Proof}
For the object $O$
let $|i\rangle$ and $|j\rangle$ a pair of eigenvectors
of $\gamma$ maximizing the expression (\ref{mdd})  and 
for the object $\tl{O}$ let
$|l\rangle$ and $|k\rangle$ be such a maximizing  pair of  eigenvectors
 of $\tilde{\gamma}$.
Then we have
\[
f_{|i\rangle\otimes |l\rangle ,|j\rangle \otimes |k\rangle}=
f_{|i\rangle ,|j\rangle}+f_{|l\rangle,|k\rangle}.
\]

In case that the signs of the two terms on the right hand side do not agree, we
can change it by exchanging $|l\rangle$ and $|k\rangle$ due to the
antisymmetry of $f$.

Assume that $\rho$ commutes with $\gamma$.
Let $Q_i$ be the spectral projections of $\gamma\otimes \tl{\gamma}$
with the corresponding eigenvalues $\mu_i$.
Write $Q_i$ as 
\[
Q_i=\oplus_l   ( P^i_l\otimes \tl{P}^i_l)
\]
where $P^i_l$ and $\tl{P}^i_l$ are spectral projections of $\gamma$
and $\tl{\gamma}$ with  eigenvalues $\lambda^i_l$ and $\tl{\lambda}^i_l$
 (respectively) such that $\lambda^i_l \tl{\lambda}^i_l=\mu_i$.
For any matrix let $a$ denote its pseudoinverse by $a^{-1}$.
Due to Lemma \ref{Umformulierungmdd}
 the maximal diagonal deviation can
be written in the form
\begin{eqnarray*}
D(O\times \tl{O})&=&\\ \max_{i,j}\Big\{|\ln \|Q_i (\rho &\otimes&
\tl{\rho}) Q_i\|+
\ln \|(Q_j (\rho\otimes \tl{\rho}) Q_j)^{-1}\|\\&-&
 \ln (\mu_i)+ \ln (\mu_j)|\Big\}.
\end{eqnarray*}

From $[\rho,\gamma]=0$ we conclude
 $P^i_l\rho P^i_m=0$ for $l\neq m$. 
Hence we have:
\begin{eqnarray*}
D(O\times \tl{O})&=&\max_{i,j}\Big\{| \ln \|\oplus_l ((P_l^i\rho P_l^i)\otimes (\tl{P}^i_l
\tl{\rho}\tl{P}^i_l))\|\\
&&+\ln \|\oplus_l ((P_l^j\rho P_l^j)^{-1}\otimes (\tl{P}^j_l
\tl{\rho}\tl{P}^j_l)^{-1})\|\\
&&- \ln \mu_i+\ln \mu_j|\Big\}\\&=&
\max_{i,j}\Big\{| \ln\max_l\| (P_l^i\rho P_l^i)\otimes (\tl{P}^i_l
\tl{\rho}\tl{P}^i_l)\|\\&&+\ln \max_l \| (P_l^j\rho P_l^j)^{-1}
 \otimes (\tl{P}^j_l
\tl{\rho}\tl{P}^j_l)^{-1}\|\\
&&-\ln \mu_i + \ln \mu_j|\Big\}\\
&\leq& D(O)+D(\tl{O})
\end{eqnarray*}
\end{Proof}

Since the maximal diagonal deviation vanishes for every equilibrium object
we have:

\begin{Corollary}
The maximum diagonal deviation is stable with respect to a composition
with arbitrary systems in its equilibrium state, i.e. we have
\[
D(O\times O_e)=D(O)
\]
for every object $O$ and every equilibrium object 
$O_e:=(\tl{\gamma},\tl{\gamma})$.
\end{Corollary}

Despite the fact, that the quantity $D$ is not additive in general,
its asymptotical  
 increase for composition of a large number  $n$ of identical objects
 is of the order $n$:

\begin{Lemma}
Let $O$ be an arbitrary object. Then 
\[
\lim_{n\to\infty}\frac{D(O^n)}{n}
\]
exists.
\end{Lemma}

\begin{Proof}
Set $O:=(\rho,\gamma)$ and
 $f(n):= D(O^n)$. 
Firstly we show that
 the sequence $f(n)/n$ is bounded from above:
Due to Lemma \ref{Umformulierungmdd} and the triangle inequality
one has
\begin{eqnarray*}
D(O^n)&\leq& |\ln \|\rho^{\otimes n}\|\,| +|\ln \|\rho^{-n}\|\,|\\
&+&
| \ln \|\gamma^{\otimes n}\|\,| + |\ln \|\gamma^{-n}\|\,|\\
&=& n (|\ln \|\rho\|\,| +|\ln \|\rho\|\,|+
 | \ln \|\gamma\|\,| +  |\ln \|\gamma\|\,|)
\end{eqnarray*}
Due to the superadditivity of $D$ one concludes
\[
f(lm+r)\geq l f(m)+f(r) \,\,\, \forall m,l,r \in \N.
\]
Now let $m$ be fixed.
For any $n$ define $l_n:=\lfloor (n/m)\rfloor$ and
$r_n:=n-m\,l_n$, hence $n=l_n\, m +r_n$, where
$\lfloor.\rfloor$ denotes the integer part of a real number.
We have
\[
\frac{f(n)}{n}=\frac{f(l_n\,m+r_n)}{l_n\, m+r_n}\geq \frac{l_n\, f(m)+f(r_n)}{l_n\, m+r_n}.
\] 
Since the right hand term tends to $f(m)/m$ for $n \rightarrow \infty$,
we conclude, that no accumulation point of $f(n)/n$ can be smaller
than $f(m)/m$. Because $m$ is arbitrary, $f(n)/n$ can have only one cumulation
point.
\end{Proof}

The maximal diagonal deviation can be interpreted geometrically:
For any pair $|i\rangle$ and $|j\rangle$  of states 
set  
\beq\label{pidef}
p_i:=\langle i|\rho|i\rangle \,\,\hbox{  and }\,\,
p_j:=\langle j|\rho|j\rangle.
\eeq
Then we consider the vector
\[
v_{i,j}(O)
\]
defined as in equation (\ref{vDef}) and note, that $D$
is given by maximizing the length of the projection of the vector 
$v_{i,j}(O)$
on the straight line $y=-x/\beta$.

The quantity $D(O)$ shows an interesting symmetry which will turn out to be
important in the theory of heating and cooling. This can be seen
by the introducing the following terminology:

\begin{Definition}
For any inverse temperature $\beta_1$ we call 
\[
\beta_2:= 2 \beta -\beta_1
\]
its {\bf complementary inverse temperature}
relative to the reference temperature $\beta$.
\end{Definition}

One checks easily that two  qubits (with the same energy gap)
 having inverse temperatures
$\beta_1$ and $\beta_2$ have the same maximal diagonal deviation from 
their equilibrium state.
Furthermore we find that
for any temperature, its complementary value is available by coupling
the considered system to an equilibrium object: 

Assume we have a pair $|i\rangle$ and $|j\rangle$
 of eigenstates of the Hamiltonian with relative inverse temperature
$\beta_{|i\rangle,|j\rangle}$. Take a qubit with energy difference
$E:=2(E_i-E_j)$ with inverse temperature $\beta$.
Then we find 
\[
\beta_{|i\rangle\otimes |0\rangle,|j\rangle\otimes |1\rangle}=
2\beta-\beta_{|i\rangle,|j\rangle},
\] 
i.e., the complementary temperature is available for 
a pair of states with the same energy gap as the original one.

As a consequence we see, that if very high relative temperatures are
inherent in an object, then very low temperatures are inherent in the
composition with an equilibrium object.
Furthermore, lower and upper limit temperatures of any object 
$O$ are determined
by $D(O)$. In order to state this more precisely we define:

\vspace{0.3cm}
\noindent
\begin{Definition*} {\bf 2 (new)}
{\it
Let $Q$ be a qubit in any diagonal state.
Let $O$ be an arbitrary object. We say, $O$ can be used for cooling
or heating  $Q$, respectively,
if there is an equilibrium object $O_e$ such that there is
an allowed transformation on $O\times O_e\times Q$ decreasing
  the occupation probability
for the upper or lower  state, respectively.}
\end{Definition*}
\vspace{0.3cm}

From now on we will use 
this terminology (in contrast to Definition \ref{tauglichDef}) and obtain:

\begin{Theorem}\label{aplmdd}
Let $Q$ be a qubit in any diagonal state.
An  object $O$ can be used for cooling {\em and} heating $Q$ if and
only if
\[
D(O)> D(Q).
\]
In the case that
\[
D(O)\leq D(Q)
\]
the resource $O$ is {\bf worthless} in the sense, that it
 can only be used for cooling if $Q$ is hotter than the equilibrium state
and it can be used for heating if $Q$ is colder than the equilibrium state.
\end{Theorem}

\begin{Proof}
Let $Q:=(diag(s,r),diag(t,v))$, where $r$ is the occupation probability for
the upper state.
Following Theorem \ref{Theorelativ} we know that $O\times O_e$ can be used
for cooling if and only if
there is a pair of states $|i\rangle$ and $|j\rangle$ 
in the composed system such that
\[
p_i/p_j < r/s,
\]
where we have taken the abbreviations given by
equation (\ref{pidef})
and $E_i-E_j=E$ if $E_i,E_j$ are the corresponding energies
and $E>0$ is the energy gap of the qubit.
Assume  the qubit to be colder than the environment. Then
\[
-D(Q)= \ln (r/s)+\beta E.
\]
By definition of $D$ we have
\[
-D(O\times O_e)+\beta E \leq \ln (p_i/p_j) \leq D(O\times O_e)+\beta E.
\]
Assume $D(O)\leq D(Q)$.
Using $D(O\times O_e)=D(O)$ and equation (\ref{mdd}) one concludes
\[
\ln(r/s) \leq \ln (p_i/p_j),
\] 
hence cooling is impossible.
Hence we see, that $D(O)>D(Q)$ is necessary for cooling an already
 cold qubit.
Similarly one shows, that this condition is necessary for heating a hot one.

Assume $D(O)>D(Q)$.
Choose a pair of states $|i\rangle$ and $|j\rangle$ of the object $O$
 such that
\[
D(O)=|\ln(p_i/p_j)+\beta (E_i-E_j)|.
\]
Due to the antisymmetry of the right hand term 
with respect to $i$ and $j$ we can even assume
\[
D(O)=\ln(p_i/p_j)+\beta (E_i-E_j)
\]
without loss of generality.
By definition of $D(Q)$ we have
\[
\ln (p_i/p_j) +\beta (E_i-E_j) > |\ln (r/s)+\beta E|.
\]
We conclude
\[
\ln (p_i/p_j) +\beta (E_i-E_j) > -\ln (r/s) -\beta E
\] 
and
\[
\frac{\ln (p_i/p_j) +\beta (E_i-E_j+E)}{E} > \frac{- \ln (r/s)}{E}.
\]
Now we take an ancilla qubit with the energies  $\tl{E}:=E_i-E_j+E$
and $0$ for the states $|1\rangle$ and $|0\rangle$.
Note that here $|1\rangle$ need not be the upper state since we do not assume
$\tl{E}>0$.
In the composition of $O$ with the ancilla qubit the pair of states
\[
|i\rangle \otimes |1\rangle \,\, \hbox{ and }\,\, |j\rangle\otimes |0\rangle
\]
have the energy gap $E$ and for the  relative inverse temperature of this pair
we conclude:
\[
\beta_{|i\rangle \otimes |1\rangle , |j\rangle\otimes |0\rangle}=
\frac{\ln (p_i/p_j)+\beta (E_i-E_j+E)}{E}> \frac{-\ln (r/s)}{E}.
\]
Hence this pair can be taken for cooling due to Theorem \ref{Theorelativ}.
In a similar way one can conclude that the object $O$ can serve
for heating.
\end{Proof}

The statement of Theorem \ref{aplmdd} can be reformulated as follows:
The lower and upper limit temperatures of an object $O$
 are given by the temperatures
of the two diagonal states $\rho_{1,2}:=diag(s_{1,2},r_{1,2})$
 of the qubit $Q:=(diag(s_{1,2},r_{1,2}),diag(t,v))$
with the property $D(Q)=D(O)$.

To avoid false conclusions at this point we emphasize
that in general
a {\em single} copy of an object $O$ is not sufficient for cooling
or heating the qubit down or up to the limit temperatures 
if the latter has the reference temperature initially.
The limit temperatures can only be approached by running 
an infinite number of stages of the same
cooling or heating procedure. This requires
an infinite number of copies of the object $O$ since the resource
has to be refreshed in each stage.  

This observation leads to another natural question:
Given any object $O$, what is the lowest temperature  
of the qubit which can be prepared by using one single copy of
the resource $O$ if the initial state of the qubit has
the reference temperature. 
One can formulate this problem more generally:
Assume we have an object $O$ and 
any other system being in its equilibrium state
$\tl{\gamma}$ initially. Which states $\tl{\gamma}$ of the latter system
can be prepared by coupling it to the object $O$ and arbitrary ancilla
equilibrium objects? With other words: Which objects $\tl{O}:=(\tl{\rho},\tl{\gamma})$ can be obtained with the help of the resources $O$? 
This  leads straightforwardly to a relation which is like
a quasi-ordering on the
set of objects, which we shall call the
{\it conversion order}:

\begin{Definition}
We say `the object $\tl{O}:=(\tl{\rho},\tl{\gamma})$ can be obtained by
using the resource $O:=(\rho,\gamma)$', formally written as
\[
O\geq \tl{O},
\]
if there exists a sequence of equilibrium objects 
$O_{e,n}:=(\hat{\gamma}_n,\hat{\gamma}_n)$ and a sequence of 
allowed transformations $u_n$ on
\[
O\times O_{e,n}\times \tl{O}
\]
such that
\[
\lim_{n\to\infty} tr_{12}(u_n (\rho\otimes \hat{\gamma}_n \otimes \tl{\gamma})u_n^*)=\tl{\rho},
\]
where $tr_{12}$ denotes the partial trace over the left most and the middle
component in the tensor product.
\end{Definition}

It is easy to give the following necessary condition
for $O\geq \tl{O}$:

\begin{Theorem}\label{Hauptsatzpartial}
Let $\cH$ and $\tl{\cH}$ be the Hilbert spaces
corresponding  to   the objects $O:=(\rho,\gamma)$ and
 $\tl{O}:=(\tl{\rho},\tl{\gamma})$, respectively.
If $O\geq \tl{O}$ then  there is a completely positive trace preserving
 map $G$
from the set of density matrices on $\cH$ to the set of density matrices
on $\tl{\cH}$ satisfying 
\beq\label{GBed1}
G\rho =\tl{\rho} \,\,\, \hbox{ and }\,\,\,
G\gamma=\tl{\gamma}
\eeq
as well as the covariance condition
\beq\label{GBed2}
[\tl{H},G(.)]=G([H,.]),
\eeq
where $H$ and $\tl{H}$  are Hamiltonians corresponding to the
equilibrium states $\gamma$ and $\tl{\gamma}$, respectively.
\end{Theorem}

\begin{Proof}
For every equilibrium object $O_{e,n}:=(\hat{\gamma}_n, \hat{\gamma}_n)$ 
and every allowed transformation $u_n$ on
$O\times O_{e,n}\times \tl{O}$ we define
\[
G_n(\sigma):= tr_{12}( u_n (\sigma \otimes \hat{\gamma}_n\otimes \tl{\gamma}) u_n^*)
\]
for every density matrix $\sigma$ on $\cH$.
Every
$G_n$ is a completely positive trace preserving map satisfying 
$G_n(\gamma)=\tl{\gamma}$ since conjugation by $u_n$ preserves
the equilibrium state of the total system.
Since the set of completely positive trace preserving maps 
for given spaces $\cH$ and $\tl{\cH}$ is compact,
the sequence $G_n$ has a convergent subsequence. Let
$G$ denote its limit point. Obviously
we have $G(\rho)=\tl{\rho}$ and $G(\gamma)=\tl{\gamma}$.
The covariance condition $[\tl{H},G(.)]=G([H,.])$ follows easily
from the fact that the allowed transformation commutes with the 
free evolution of the total system and preserves the equilibrium states
in every tensor component.
\end{Proof}

In the following we will try to work out the conversion order
as explicitly as possible.
We start by doing this for quasi-classical objects.
In this case the
 quasi-ordering can be given explicitly:

\begin{Theorem}\label{stochasticTheo}
Let $O:=(p,g)$ and $\tilde{O}:=(\tl{p},\tl{g})$ be quasi-classical objects.
Then 
\[
O\geq \tilde{O}
\]
if and only if there is a stochastic matrix $A$ such that
\beq\label{stochastic}
Ap=\tilde{p} \,\,\,\hbox{ and } \,\,\,
Ag=\tilde{g}.
\eeq
\end{Theorem}

\begin{Proof}
Assume $O\geq \tilde{O}$.
Let  $\rho,\gamma,\tl{\rho},\tl{\gamma}$ be the density matrices with diagonal
entries $p,g,\tl{p}$, and $\tl{g}$, respectively.
Assume $p,g \in \R^l$ and $\tl{p},\tl{g} \in \R^{\tl{l}}$.
For any density matrix $\sigma$ acting on $\C^{\tl{l}}$ we define the
vector $q(\sigma)\in \R^{\tl{l}}$ by the diagonal  of $\sigma$. 
For every $i\leq l$ define the density matrix
\[ 
e_i:=diag (0,\dots,0,1,0,\dots,0)
\]
where the entry `$1$' is on position $i$.
Define a stochastic $\tl{l}\times l$-matrix $A$  by
\[
Ar=q(G(\sum_i r_i e_i)),
\]
where $r_i$ is the $i$-th component of an arbitrary vector $r\in \R^{l}$.
Obviously, both equations in (\ref{stochastic}) are fulfilled.

Assume there is a stochastic matrix $A$ such that 
$Ap=\tilde{p}$ and $Ag=\tilde{g}$.

For every $n\in \N$ choose the environment
\[
O_{e,n}:=(g^{\otimes n} \otimes \tilde{g}^{\otimes n},g^{\otimes n} \otimes 
\tilde{g}^{\otimes n}). 
\]
Assume $p\in \R^l$ and $\tl{p}\in \R^{\tl{l}}$.
 Hence the  pure states of the systems described by $g$ and $\tilde{g}$
can be named  by the symbols
$1,\dots,l$ and the symbols $1,\dots,\tilde{l}$, respectively.

Let $\cS_n$ be the set of
pure  states in the composed system described by the equilibrium state
\[
g\otimes g^{\otimes n}\otimes \tilde{g}^{\otimes n} \otimes 
\tilde{g}.
\]
Every element of $\cS_n$ is
  characterized by a word of
length $n+1$ over the alphabet $\{1,\dots,l\}$ and
a  word of length $n+1$ over the alphabet
$\{1,\dots, \tilde{l}\}$. 
In the following, only four attributes of these
 word pairs are relevant:
\begin{enumerate}
\item 
the first symbol of the first word, denoted by $j$.
\item 
the numbers of occurrences of the symbols $1,\dots,l$ in the
first word, denoted by
$r_1,\dots,r_l$, or simply by the vector $r\in \N^{\,l}$ with $\sum_i r_i=n+1$.
\item
the numbers of occurrences of symbols $1,\dots,\tilde{l}$ in the second word,
denoted by $s_1,\dots, s_{\tilde{l}}$ or the vector $s\in \N^{\,\tilde{l}}$
with $\sum_i s_i=n+1$.
\item
the last symbol of the second word, denoted by $x$.
\end{enumerate}

Hence we assign the 4-tuple $(j,r,s,x)$ to every pair of words.
Now let $n,r,s$ be fixed.
Note that all the states with a common vector $r$ and $s$ have the same 
energy.
We write $\{(j,r,s,.)\}$ for the cylindric set of states
having  $j,r,s$ as the first three attributes.
Their numbers of elements are given by a product
of two  multinomial coefficients
\beq\label{bj}
b_j:=\frac{r_j\,\,n!}{\prod_{i\leq l} (r_i)!}\frac{(n+1)!}{ 
\prod_{i\leq \tl{l}} (s_i)!}
\eeq
Accordingly, write $\{(.,r,s,x)\}$ for the set of states 
with $r,s,x$ as the last three attributes.
Their numbers of elements are given by
\beq\label{cx}
c_x:=\frac{(n+1)!}{\prod_{i\leq l} (r_i)!}
\frac{s_x\,\, n!}{ \prod_{i\leq \tl{l}} (s_i)!}
\eeq
Note that  these sets depend on the number $n$, i.e., the size
 of the environment, although  we do not indicate this explicitly
 by indices.

Let $a_{xj}$ with $j\leq l,x\leq \tl{l}$ be the entries
 of the matrix $A$.
Now we define for each $x$ the $l$  numbers
\[
m_{xj}:=\min \{c_x-\sum_{i<j} m_{xi}, \lfloor  a_{xj}b_j \rfloor\},
\]
where $\lfloor.\rfloor$ denotes the integer part of a real number.

For each $j$ choose $\tl{l}$ disjoint sets  $M_{xj}\subset \{(j,r,s,.)\}$
 with $m_{xj}$ elements.
This is possible since 
\[
\sum_x m_{xj} \leq \sum_x a_{xj} b_j=b_j.
\]
Note that we do not indicate explicitly that the numbers 
$b_j,c_x, m_{xj}$ as well as the sets $M_{xj}$ depend on $(r,s)$.
 Choose an injective map 
\[
\hat{\pi}_{r,s}: \cup_{x,j} M_{xj} \rightarrow \{(.,r,s,.)\}
\]
such that
\[
\hat{\pi}_{r,s}(M_{xj}) \subset \{(.,r,s,x)\}.
\]
This is possible since $\sum_j m_{xj} \leq c_x$.
Extend $\hat{\pi}_{r,s}$ to a bijection
\[
\pi_{r,s}: \{(.,r,s,.)\}\rightarrow \{(.,r,s,.)\}.
\]
Now perform such a transformation $\pi_{r,s}$ on  every set  
$\{(.,r,s,.)\}\subset \cS_n$. For every $n$, this defines a bijection
\[
\pi_n: \cS_n\rightarrow \cS_n.
\]

Let $P_n$ be the probability measure on $\cS_n$ defined by the
composed system's initial state
\[
p\otimes g^{\otimes n}\otimes \tl{g}^{\otimes n}\otimes \tl{g}.
\]
Let $\tl{P}_n$ be the image of $P_n$ under the transformation $\pi_n$, i.e.,
\[
\tl{P}_n:=P_n \circ \pi_n^{-1}.
\]
Let $\T_n\subset \{ (r,s) \in \N^{\,l}\times \N^{\,\tl{l}}\,|\,
\sum r_i=n+1,\,\, \sum s_i=n+1\} $ be a such that
\begin{eqnarray*}
&&\lim_{n\to\infty}\sum_{(r,s)\in \T_n} \tl{P}_n(\{(.,r,s,.)\})=\\
&&\lim_{n\to\infty} \sum_{(r,s)\in \T_n} P_n(\{(.,r,s,.)\}=1
\end{eqnarray*}
and
\[
\lim_{n\to\infty} \max_{(r,s) \in \T_n} \{ \|\frac{r}{n}-g\|+\|\frac{s}{n}-\tl{g}\|\} 
= 0.
\]
This is possible due to  the law of large numbers,
since the words with 
\[
\|\frac{r}{n}-g\|\approx 0 \,\,\hbox{ and } \,\, \|\frac{s}{n}-\tl{g}\|\approx 0
\]
 are {\em typical} (c.f.\cite{CT}).

Now we have to show, that asymptotically  
 the probabilities of the symbols $1,\dots,\tl{l}$
in the right most component of the system
are changed 
from $\tl{g}_1,\dots,\tl{g}_{\tl{l}}$ to
$\tl{p}_1,\dots,\tl{p}_{\tl{l}}$ by the permutations $\pi_n$, i.e.,
we must show
\[
\sum_{r,s} \tl{P}_n (\{(.,r,s,x)\})\rightarrow \tl{p}_x.
\]
We do this by proving
\[
\max_{(r,s)\in \T_n} |\frac{\tl{P}_n(\{(.,r,s,x)\})}{\tl{P}_n(\{(.,r,s,v)\})}
-\frac{\tl{p}_x}{\tl{p}_v}|\rightarrow 0.
\]
With respect to the initial probability measure $P_n$ every word pair
with attributes $(j,r,s,x)$ has the probability
\beq\label{wj}
w_j:=\frac{p_j}{g_j} \prod_{i\leq l} g_i^{r_i} \prod_{i\leq \tl{l}}
 \tl{g}_i^{s_i}.
\eeq
If for every $n$ our attention is only restricted to those
 vector pairs  $(r,s)$ which are elements of  
 $\T_n$, we have the following asymptotic statements
as $n$ goes to infinity:

\begin{enumerate}
\item
The quotients $c_x/b_j$ tend to $\tl{g}_x/g_j$ 
and $b_j/b_i$ tend to $g_j/g_i$
due to equations (\ref{bj}) and (\ref{cx}).

\item Therefore $m_{xj}/b_j\rightarrow a_{xj}$. This follows from
1 by induction over $j$ because $\sum_j a_{xj}g_j=\tl{g}_x$.

\item The set $\{(j,r,s,.)\}$ is more and more exhausted by
$\cup_x M_{xj}$ in the sense that the number of elements of its complement
becomes negligible compared to the number of elements of 
$\{(j,r,s,.)\}$.
This shows that the total probability of the complement becomes
 irrelevant, since all its elements have the same probability.
\end{enumerate}

We conclude:
\begin{eqnarray*}
\lim_{n\to\infty} \max_{(r,s) \in \cT_n}&& \frac{\tl{P}_n(\{(.,r,s,x)\})}{\tl{P}_n
(\{(.,r,s,v)\})}\\
=\lim_{n\to\infty}\max_{(r,s)\in \cT_n}&& \frac{\sum_j m_{xj}w_j}{\sum_j m_{vj}w_j}\\
&=&\frac{\sum_j a_{xj} g_j w_j}{\sum_j a_{vj} g_j w_j}=\frac{\sum_j a_{xj}p_j}{\sum_x a_{vj} p_j}=\frac{\tl{p}_x}{\tl{p}_v}.
\end{eqnarray*}

For reasons of convenience we
 dropped the index $n$ for $r,s,m_{xj},b_j$.

The reason for the first equality is given by statement 3. The second one
is proven by the statements 1 and 2. The third equality is due to
equation (\ref{wj}) and the last one by assumption. 

The statements 1-3 reflect the following idea behind our construction:
The part $a_{xj}$ of the elements in $\{(j,r,s,.)\}$ is mapped
onto an element in $\{(.,r,s,x)\}$. Since the ratios of the sizes of these
  sets behave asymptotically as $g_j:\tl{g}_x$,
the condition $\sum_j a_{xj}g_j=\tl{g}_x$ guarantees that
 such a map
can be constructed as a bijective one.
For typical $(r,s)$,
the numbers of elements in $\{(1,r,s,.)\},\dots,\{(l,r,s,.)\}$
 are related to each other
by $g_1,\dots, g_l$ and the probabilities of single elements
in $\{(1,r,s,.)\},\dots,\{(l,r,s,.)\}$ are related by $p_1/g_1,\dots,p_l/g_l$.
The total probability of the set $\{(.,.,.,x)\}$ after having performed
the transformation is therefore
given by $\sum_j a_{xj}g_j p_j/g_j=\sum_j a_{xj} p_j=\tl{p}_x$. 
\end{Proof}

Loosely speaking, we have shown, that any stochastic matrix,
which maps an equilibrium state of the first system on the equilibrium
state of the second one, can be carried out by an energy conserving process
provided that any ancilla system being in its equilibrium state can be used.

Note that Theorem \ref{stochasticTheo} shows a symmetry with respect to an exchange
of the actual probability distribution $p$  and the equilibrium distribution
$g$:

\begin{Corollary}
We have:
\[
(p,g)\geq (\tilde{p},\tilde{g})
\]
if and only if
\[
(g,p)\geq (\tilde{g},\tilde{p}).
\]
\end{Corollary}
The physical consequences of this symmetry are by no means obvious.
Its investigation has to be left to the future.

Due to the convexity of the set of stochastic matrices 
we conclude:

\begin{Corollary}
Let $O$  be an arbitrary object.
Let $\hat{O}:=(\hat{p},g)$ and $\tl{O}:=(\tl{p},g)$ two identical systems
being in  different  states.
Then $O\geq \hat{O}$ and $O\geq \tl{O}$ implies
\[
O\geq (\lambda \hat{p}+ (1-\lambda) \tl{p}, g))
\]
for every $0\leq \lambda \leq 1$.
\end{Corollary} 

Obviously, it is not satisfactory to restrict the analysis to the
quasi-classical case. 
 Fortunately, there are many cases where the investigation of the conversion
order can be reduced to the conversion order on the quasi-classical 
objects and then Theorem \ref{stochasticTheo} is used for proving considerably
more general theorems. For that purpose we need a definition and a technical lemma:

\begin{Definition}
Let $O:=(\rho,\gamma)$ be an arbitrary object and $B$ be a 
basis diagonalizing $\gamma$. Let $p$ and $g$ be the vectors
given by the diagonal entries of $\rho$ and $\gamma$, respectively.
Then we define the corresponding quasi-classical object
\[
C_B(O):=(p,g)
\]
with respect to the basis $B$.
\end{Definition}

\begin{Lemma}\label{reductionquasi}
Let $O:=(\rho,\gamma)$ be an arbitrary object.
For any basis $B$ diagonalizing $\gamma$ we have
\[
O\geq C_B(O).
\]
\end{Lemma}

\begin{Proof}
Let $B$ be given by $B:=\{|1\rangle,\dots,|l\rangle\}$.
Let $\sigma$ be the maximally mixed state in $l$ dimensions.
Take the equilibrium object $O_e:=(\sigma,\sigma)$.
With the help of $O_e$ we can obtain $C_B(O)$ by using the resources
$O$: Take the initial state 
 $\rho\otimes \sigma \otimes \gamma$
of the tripartite  system $\gamma \otimes \sigma\otimes \gamma$
  and perform the transposition
\[
|i\rangle \otimes |j\rangle \otimes |k\rangle \,\,\, \leftrightarrow \,\,\, 
|k\rangle \otimes |j\oplus i \rangle \otimes |i\rangle,
\]
where $\oplus $ denotes the addition modulo $l$.
This transformation is energy conserving since the equilibrium object is
degenerated and the other systems have identical level structure.
Obviously the transformation transfers the diagonal entries
of $\rho$ to the other identical system and destroys the coherence since
the coupling to the degenerated ancilla system acts like a measurement.
\end{Proof}
  
We are now able to draw some important conclusions:

\begin{Theorem}\label{Umkehrung}(partial converse of Theorem \ref{Hauptsatzpartial})
If at least one of the two objects $O:=(\rho,\gamma)$
and   $\tl{O}:=(\tl{\rho},\tl{\gamma})$
is quasi-classical, i.e.,
\[
[\rho,\gamma]=0 \,\,\, \hbox{  {\em or} }\,\,\, [\tl{\rho},\tl{\gamma}]=0
\]
the following equivalence holds:
\[
O\geq \tl{O}
\]
if and only if there is a completely positive trace preserving map fulfilling
equations (\ref{GBed1}) and (\ref{GBed2}).
\end{Theorem}

\begin{Proof}
Let $[\rho,\gamma]=0$. Then we have $[\rho,H]=0$ for
the corresponding Hamiltonian. Take $G$ fulfilling the equations
(\ref{GBed1}) and (\ref{GBed2})
of Theorem \ref{Hauptsatzpartial}.
Then we have:
\[
0=G([H,\rho])=[\tl{H},G(\rho)]=[\tl{H},\tl{\rho}].
\]
Hence $[\tl{\rho},\tl{\gamma}]=0$.
Hence it is sufficient to show the statement
 for the case $[\tl{\rho},\tl{\gamma}]=0$: 

Let $Q_i$ and $\tl{Q}_i$ be the spectral projections of
$\gamma$ and $\tl{\gamma}$, respectively.
Then $P(\sigma):=\sum_i Q_i \rho Q_i$
and $\tl{P}(\tl{\sigma}):=\tl{Q}_i \tl{\sigma} \tl{Q}_i$
 project
any arbitrary density matrix $\sigma$ and $\tl{\sigma}$ on its time average
 with respect to  the evolution generated by $H$ and $\tl{H}$, respectively.
Due to the covariance condition of $G$ we conclude
\[
G(P(\rho))=\tl{P}(G(\rho))=\tl{P}(\tl{\rho})=\tl{\rho}.
\]
Without loss of generality we can assume that
$P(\rho),\gamma, \tl{\rho},\tl{\gamma}$ are diagonal since 
$[P(\rho),\gamma]=0$.
For any density matrix $\sigma$ acting on $\C^m$ with arbitrary $m$ let
 $R(\sigma)$ be the density matrix obtained by cancelling the
off-diagonal entries.

We define
\[
G':=R\circ G \circ R.
\]
Due to $R(P(\rho))=P(\rho)$ and $R (\tl{\rho})=\tl{\rho}$
we see that $G'$ satisfies the equations (\ref{GBed1}) and (\ref{GBed2})
as well. Since $G'$ defines a map from
 diagonal matrices on diagonal ones it can be described by a stochastic 
matrix. Therefore we can apply Theorem \ref{stochasticTheo}
 to show that
\[
(R(\rho),\gamma)\geq (\tl{\rho},\tl{\gamma})
\]
by taking the canonical basis of $\C^l$ as $B$.
 Lemma \ref{reductionquasi} completes the proof due to the transitivity 
of the conversion order.
\end{Proof}

For $[\tl{\rho},\tl{\gamma}]=0$ 
the conversion order can be reduced to the quasi-classical case in the following
sense:

\begin{Corollary}
Let $P(\sigma)$ be  (as in the proof of Theorem \ref{Umkehrung})
the time average of any  density matrix $\sigma$.
If $[\tl{\rho},\tl{\gamma}]=0$ then the following statements are equivalent:
\begin{enumerate}
\item
$
O:=(\rho,\gamma)\geq (\tl{\rho},\tl{\gamma})=:\tl{O}
$
\item
There is a basis $B$ diagonalizing $\gamma$ and a basis $\tl{B}$ diagonalizing
$\tl{\rho}$ and $\tl{\gamma}$ simultaneously such that
\[
C_B(O)\geq C_{\tl{B}}(\tl{O}).
\]
\item For every basis $B$ diagonalizing $P(\rho)$ and $\gamma$ simultaneously
 and every basis $\tl{B}$ diagonalizing $\rho$ and $\gamma$ simultaneously
\[
C_B(O)\geq C_{\tl{B}}(\tl{O})
\]
holds.
\end{enumerate}
\end{Corollary}

\begin{Proof}
1 $\Leftrightarrow$ 3:
Like in the proof of Theorem \ref{Umkehrung} there is a stochastic matrix
mapping the diagonal entries of $R_B(\rho)$ on the diagonal entries of
$R_{\tl{B}}(\tl{\rho})$ and the same for the corresponding equilibrium states.
3 $\Rightarrow 2$: Obvious.
2 $\Rightarrow 1$: The stochastic $C$ matrix mapping the diagonal entries
of $R_B(\rho)$ onto the diagonal entries of $\tl{\rho}$
can be extended to a completely positive trace preserving map by:
\[
G:=C\circ R_B.
\]
Clearly $G$ fulfills the requirements of Theorem \ref{stochasticTheo}.
\end{Proof}

For generic pairs of objects $(\rho,\gamma)$ and $(\tl{\rho},\tl{\gamma})$
no difference of the eigenvalues of $\tl{H}$ will coincide 
with the eigenvalues of $H$.
One can show that in this case the condition $[\tl{\rho},\tl{\gamma}]=0$ is
necessary:

\begin{Lemma}
Let the energy levels of the objects $O:=(\rho,\gamma)$
 and $\tl{O}:=(\tl{\rho},\tl{\gamma})$
 be  such that no energy difference in $O$ coincides with any difference
 in $\tl{O}$.
Then $O\geq \tl{O}$ implies
\[
[\tl{\rho},\tl{\gamma}]=0.
\]
\end{Lemma}

\begin{Proof} 
Let $G$ be the completely positive map required by Theorem
\ref{Hauptsatzpartial}. Canonically, we extend $G$ to a linear map to the 
set of matrices acting on the corresponding Hilbert space.
Let $|i\rangle$ and $|j\rangle$ be eigenvectors of $H$ with eigenvalues
$E_i$ and $E_j$. Then $|i\rangle\langle j|$ is an eigenvector of
the operator $[H,.]$ with eigenvalues $E_i-E_j$.
Due to $G([H,.])=[\tl{H},G(.)]$ the density matrix
 $G(|i\rangle\langle j|)$ has to be an 
eigenvector of the superoperator
 $[\tl{H},.]$ with eigenvalues $E_i-E_j$ as well.
But there are no eigenvectors with this eigenvalue by assumption.
Hence $G(|i\rangle\langle j|)=0$. Hence every density matrix 
in the image of $G$ commutes with $\tl{H}$ and 
$\tl{\gamma}$.
\end{Proof}

Now we will show, that the
 problem of cooling a qubit is indeed a typical application
of the conversion order. In our formal setting
we can formulate it as follows:
Cooling the qubit down to the temperature $\hat{T}$ 
means preparing the object
\[
\tl{O}:=( \left(\begin{array}{cc} \tl{p}_1 & 0\\0 & \tl{p}_2
 \end{array}\right),\left(\begin{array}{cc} \tl{g}_1 & 0\\0 & \tl{g}_2
\end{array} \right))
\]
with
\[
\tl{p}_1:=\frac{1}{1+e^{-E/(k\hat{T})}} 
\,\,\hbox{ and }\,\,
\tl{p}_2=1-\tl{p}_1
\]
as well as
\[
\tl{g}_1=\frac{1}{1+e^{-E/(kT)}} 
\,\,\hbox{ and }\,\,
\tl{g}_2=1-\tl{g}_1.
\]
 
For given resources $O$ it seems hard to decide whether
there is a completely positive map as specified by Theorem \ref{Umkehrung}.
Fortunately the problem turns out to be equivalent to a well-known 
 problem of testing hypotheses: If one wants to decide whether a given state is the state $\rho$ or 
the state $\gamma$ one has to construct a measurement such that the 
measurement outcome tells whether $\rho$ or $\gamma$ is more likely. 
Such a {\em decision rule} can be described by a {\em positive operator valued
measure} $({\cal E}_\rho,{\cal E}_\gamma)$ where ${\cal E}_\rho$ and 
${\cal E}_\gamma$ are 
positive operators on the resource's Hilbert space with ${\cal E}_\rho+{\cal E}_\gamma=1$.
Then the risk of the {\em error of the first kind}, i.e., 
the risk of deciding $\rho$ if  $\gamma$ is actual, is given by
\[
F_1:=tr(\gamma {\cal E}_\rho)
\]
and the risk of the error of the second kind is given by
\[
F_2:=tr(\rho {\cal E}_\gamma).
\]

If we want to distinguish between
the qubit state with temperature $T$ and the state with temperature $\hat{T}$,
a straightforward  decision rule would be given by measuring whether
the system is in its upper or in its lower state.
In the first case we will decide to have the higher temperature $T$, else
we decide for $\hat{T}$.
This would be a decision rule with the error probabilities
\beq\label{errorTT}
F_1=\frac{e^{-E/(k\hat{T})}}{1+e^{-E/(k\hat{T})}}
\,\,\hbox{ and }\,\,
F_2=\frac{1}{1+e^{-E/(kT)}}  .
\eeq
If the cold qubit has been prepared by using the resources $(\rho,\gamma)$
one can define a decision rule for the distinction between $\rho$ and $\gamma$
with the same error probabilities by
\[
({\cal E}_{\tl{\rho}}\circ G, {\cal E}_{\tl{\gamma}}\circ G),
\]
where $(\E_{\tl{\rho}},\E_{\tl{\gamma}})$
is the 
decision rule described above  and $G$ is a completely positive trace preserving
map with the required properties. 
Hence the resources $(\rho,\gamma)$ can only be used for cooling
the qubit down to the temperature $\tl{T}$ if there is a decision rule
\[
(\E_\rho,\E_\gamma)
\]
with the error probabilities given by equations (\ref{errorTT}).
As one of the main results of our theory, it turns out that 
this condition is even sufficient:

\begin{Theorem}\label{Entscheidung}
The resource $(\rho,\gamma)$ can be used for cooling the qubit
down to the temperature $\hat{T}$ if and only if
there is a decision rule $({\cal E}_\rho,{\cal E}_\gamma)$ with
 $[{\cal E}_\gamma,\gamma]=0$ such 
that the errors are given by
\[
F_1=\frac{e^{-E/(k\hat{T})}}{1+e^{-E/(k\hat{T})}}
\]
\[
F_2=\frac{1}{1+e^{-E/(kT)}}  .
\]
\end{Theorem}

\begin{Proof}
That the condition is necessary has already been explained above.

The other direction can be seen as follows:
Define a map from the density matrices on the  Hilbert space of $O$ by:
\[
G(\sigma):=\left(\begin{array}{cc} tr({\cal E}_\gamma \sigma) & 0 
\\ 0 &tr({\cal E}_\rho \sigma) \end{array}\right)
\]
The map $G$ is completely positive since every positive map
with a commutative image is completely positive.
Furthermore it fulfills the requirements of Theorem \ref{Umkehrung} (Note that
we have $[\E_\gamma,\gamma]=0$ by assumption).
\end{Proof}

One may question the practical importance of the
converse direction 
which  states that a cooling procedure is possible if
the conditions of Theorem \ref{Entscheidung} are satisfied,
since we used rather sophisticated unitary transformation in the proof
of Theorem \ref{Umkehrung}. However, it is not clear
whether a more suitable environment  (e.g. an infinite dimensional one)
  might allow  optimal transformations which are much more natural.
Furthermore it is an important insight that  it is not possible to derive
any tighter bounds for the resources within our setup.

If any resource object $O:=(\rho,\gamma)$ is given and the criterion
of Theorem \ref{Entscheidung} tells that $O$ is not sufficient
for obtaining the demanded temperature, it is a natural question
whether sufficient cooling is enabled by using many copies
of the object $O$. Therefore one would ask for the least $n$ such that
the resource object $O^n:=(\rho^{\otimes n},\gamma^{\otimes n})$
is sufficient for preparing a qubit with temperature $\tl{T}$.
Using Theorem \ref{Entscheidung}, this is the  question of the {\em increase} 
of the {\em distinguishability} between the states
$\rho^{\otimes n}$ and $\gamma^{\otimes n}$ (see \cite{Fu}).
 However, it is important
to note that the condition
$[{\cal E}_\gamma,H]=0$ in Theorem \ref{Umkehrung} differentiates 
the problem from the 
usual information theoretic questions.
--- Note that there can be an abundance of basis
diagonalizing $\rho^{\otimes n}$ and $\gamma^{\otimes n}$
simultaneously. Therefore the application of Theorem \ref{Entscheidung}
is by no means easy!
We will restrict our attention to the quasi-classical case,
where we can use essentially Stein's Lemma \cite{Bl} 
of classical information theory:

\begin{Theorem}\label{lowtemp}
For a quasi-classical object $O:=(p,g)$ define its Kullback-Leibler
Relative Information as
\[
S(g\,\|\,p):=\sum_i g_i\ln \frac{g_i}{p_i}.
\]
We consider the situation where the
 $n$-fold copy of this resources $O^n:=(p^{\otimes n},g^{\otimes n})$
is used for cooling a two-level system with energy gap $E$. Let  $T_n$
denote the lowest obtainable temperature. Then we have:
\[\lim_{n\to \infty}n\,k\,T_n=E\,S(g\,\|\,p),
\]
where $k$ is Boltzmann's constant.
\end{Theorem}

\begin{Proof}
Let $\tl{g}:=(\tl{g}_1,\tl{g}_2)$ be the equilibrium  state of the qubit.
Let ${\bf i}:=(i_1,\dots,i_n)\in \{1,\dots,l\}^n$ 
be a pure state in the $n$-fold copy of the system.
Then, instead of working with positive operator valued measurements, we can 
specify the decision rule by the conditional  probabilities 
\[
w(1|{\bf i})\,\, \hbox{ and }\,\, w(2|{\bf i})=1-w(1|{\bf i}) 
\]
describing the probability for deciding $g$ or $p$ (respectively)
 when ${\bf i}$ is measured.
The corresponding error probabilities  are given by
\[
F_1=\sum_{{\bf i}} w(2|{\bf i})\, g^{\otimes n}({\bf i})
\]
and
\[
F_2=\sum_{{\bf i}} w(1|{\bf i})\, p^{\otimes n}({\bf i}),
\] 
where we consider the vectors $p^{\otimes n}$ and $g^{\otimes n}$ as
probability measures on $\{1,\dots,l\}^n$ in a straightforward way.
Now the proof goes in strong analogy to the proof of Theorem 4.4.4
in \cite{Bl} with the difference that we have a stochastic decision rule, not
a deterministic one.
For any $\epsilon >0$ define the set $B_\epsilon \subset \{1,\dots,l\}^n$ by
\[
B_\epsilon:=\{{\bf i} |\, S(g\,\|\,p)-\epsilon < \frac{1}{n} \sum_j
\ln(g_{i_j}/p_{i_j}) < S(g\,\|\,p)+\epsilon \}
\]
Due to the law of large numbers we have: 
\[
\lim_{n\to\infty} g^{\otimes n}(B_\epsilon)=1 > \tl{g}_1
\]
Therefore, for large $n$, we can 
 define a decision rule by
\[
w(1|{\bf i}):=\frac{\tl{g}_1}{g^{\otimes n}(B_\epsilon)}\,\,\, \forall {\bf i} \in B_\epsilon
\]
and
\[
w(1|{\bf i}):=0 \,\,\,\forall {\bf i} \in \{1,\dots, l\}^n\setminus B_\epsilon.
\]
We have 
\begin{eqnarray*}
F_1&=&\sum_{{\bf i}}w(2|{\bf i})\, g^{\otimes n} ({\bf i})=1- \sum_{{\bf i}\in B_\epsilon} w(1|{\bf i})\, g^{\otimes n}({\bf i})\\&=&1-\tl{g}_1=\tl{g}_2 
\end{eqnarray*}
 as required by Theorem \ref{Umkehrung}.
Furthermore we have
\begin{eqnarray*}
F_2&=&\sum_{{\bf i} \in B_\epsilon} w(1|{\bf i})\, p^{\otimes n}({\bf i})\\
&\leq&
\sum_{{\bf i}\in B_\epsilon} w(1|{\bf i})\,g^{\otimes n}({\bf i}) 
e^{-n(S(g\,\|\,p)-\epsilon)}= \tl{g}_1 e^{-n(S(g\,\|\,p)-\epsilon)}.
\end{eqnarray*}
If the decision rule $w(.|.)$  is defined in any other way, we have
\begin{eqnarray*}
F_2&\geq &\sum_{{\bf i}\in B_\epsilon} w(1|{\bf i})\,p^{\otimes n}({\bf i})\geq
\sum_{{\bf i}\in B_\epsilon} w(1|{\bf i})\, g^{\otimes n}({\bf i})\,
e^{-n(S(g\,\|\,p)+\epsilon)}\\&=&\sum_{{\bf i}\in B_\epsilon}(1-w(2|{\bf i}))
g^{\otimes n}({\bf i})\,  e^{-n(S(g\,\|\,p)+\epsilon)}\\&\geq& (g^{\otimes n}(B_\epsilon)-F_1)\,e^{-n(S(g\,\|\,p)+\epsilon)}.
\end{eqnarray*}
With Theorem \ref{Entscheidung} we obtain:
\begin{eqnarray*}
\tl{g}_1\,e^{-n(S(g\,\|\,p)+\epsilon)}&\geq&
\frac{e^{-E/(kT_n)}}{1+e^{-E/(kT_n)}}\\&\geq& (g^{\otimes n}(B_\epsilon)-F_1)\,
 e^{-n(S(g\,\|\,p)-\epsilon)}.
\end{eqnarray*}
Since $g^{\otimes n}(B_\epsilon)$ converges to 1 and $F_1$ is constant we get:
\[
\lim_{n\to \infty} n\,k\,T_n=S(g\,\|\,p)\, E.
\]
\end{Proof}

\section{Further applications of the conversion order}

One of the big merits of the Second Law of Thermodynamics
is the restriction it puts on the efficiency of conversion
of heat to other forms of energy: A power station
working with two heat reservoirs having temperatures  $T$ and
$\tl{T}$ with $\tl{T}> T$ can never work with an efficiency above
\[\frac{\tl{T}-T}{\tl{T}}.
\]
We will show in which sense our theory puts restrictions on the efficiency
of energy conversion processes which are apparently not
given by easy conclusions from the well-known laws of thermodynamics:
Assume that we have an energy source, i.e.,
an object $(\rho,\gamma)$ such that the mean energy of the state
is above the mean energy of equilibrium, i.e., we have
\[
tr(H\rho)>tr(H \gamma ).
\]
Converting the energy to another form of energy means preparing
another object $(\tl{\rho},\tl{\gamma})$ by using $(\rho,\gamma)$ as
resource. Generically, we will not expect that it is possible
to undo the conversion, i.e.,
to prepare $(\rho,\gamma)$ by using now $(\tl{\rho},\tl{\gamma})$
as resource.
In general, for a given system $\tl{\gamma}$, we cannot expect that
there is a state $\tl{\rho}$ such that
\[
(\rho,\gamma)\geq (\tl{\rho},\tl{\gamma})\geq (\rho,\gamma).
\]
We can say: The transport of the energy to the other system is 
an irreversible process so that we cannot regain the original resources.
We will illustrate this by an  easy example with two
qubits:

Take a qubit where the upper level has a higher occupation probability
compared to the equilibrium:
\[
O:=((p_1,p_2),(g_1,g_2)) \hbox{ with } p_2 > g_2,
\]
where $p_2$ and $g_2$ denote the occupation probabilities of the upper level.
For another qubit described by the equilibrium
 probabilities $\tl{g}_1$ and $\tl{g}_2$
for the upper and lower level
let $\tl{p}_2$ be the largest probability such that
\[
(p,g)\geq (\tl{p},\tl{g}).
\]
Assume 
\[
(p,g)\geq (\tl{p},\tl{g})\geq (p,g).
\]
Then there are stochastic matrices $A$ and $B$ such that
\[
BAp=p \hbox{ and } BAg=g.
\]
Since we assume $p\neq g$ the matrix
 $BA$ must be the identity matrix. 
Therefore either $A$ and $B$ are  identity matrices or $A$ and $B$ are
transpositions exchanging the upper and lower state.
We can exclude the latter case since that would mean  that
$\tl{g}_1> \tl{g}_2$ if $g_1<g_2$ or
$\tl{g}_1< \tl{g}_2$ if $g_1> g_2$. This is not possible for any temperature.

If $B$  and $A$ are identity matrices the energy levels of both systems are compatible. In this case
it is obviously possible to transfer the energy without loss.
In all the other cases the greatest $p_2'$ such that
\[
(p,g)\geq (\tl{p},\tl{g})\geq ((1-p'_2,p'_2),g)
\]
has a value below $p_2$, i.e., we obtain a lower 
probability for the upper level compared to the initial one.
Of course  we can not apply these arguments if many copies
of these qubits are available. But even in this case we have 
the statement that energy conversion with lower loss requires
processes involving more qubits at once. Hence 
energy conversion with high efficiency
 turns out to be a matter of {\em complexity} of the
conversion process.

\section{Comparison with Landauer's principle}

To elucidate the connection of our analysis with
Landauer's principle we reformulate it within our framework.

It should be emphasized that
the formulation `The erasure of one bit of information
requires at least the dissipation of the energy $kT \ln 2$.'
 has to be read in the sense that the bit is in a totally 
unknown state, i.e., the erasure changes the probabilities
of the state $|0\rangle$ from $1/2$ to $1$.
It is straightforward to model the bit as a two-level system
being in its maximally mixed state initially.
If we assume the two-level system to be degenerated then the erasure
process fits well into our framework since the maximally mixed state
is the equilibrium state in this case.
Anyway, in the non-degenerate case it would be more complicated to 
see Landauer's principle since the two-level system may supply
the energy required for its own erasure. 
The requirement of the energy supply $kT\ln 2$ should be made more precisely:
Of course this energy cannot be supplied by 
the {\em heat} of an reservoir having the temperature $T$,
 since heat is a useless form of energy.
We rather need {\em free energy}  for driving the process.
Therefore, we need resources $(p,g)$
such that the free energy of $p$ exceeds the free energy of $g$
at by least   $kT\ln 2$.
Note that this difference of the free energies of $p$ and $g$
is given by the Kullback-Leibler Information up to Boltzmann's constant:

The free energy of any state $p$ with respect to the inverse temperature
$\beta$ is given by:
\[
F_g(p):=E_g(p)-\frac{1}{\beta}S(p), 
\]
where $S(p):=-\sum_i p_i\ln p_i$ is the entropy and
 $E_g(p)$ is the mean energy in the state $p$ (in view of the
energy level structure defined by $g$), i.e.
\[
E_g(p):=\sum_i p_i E_i.
\]
Easy calculation shows the following well-known result:
\[
F_g(p)-F_g(g)=\frac{1}{\beta}(\sum_i p_i\ln (p_i/g_i))=\frac{1}{\beta} S(p\,\|\,g).
\]
Note that here (in contrast to Theorem \ref{lowtemp})
 the relative information $S(p\,\|\,g)$ instead of $S(g\,\|\,p)$ occurs! 
Therefore, we rephrase Landauer's principle as:
`The erasure process (in the sense above) requires an
object $(p,g)$ with $S(p\,\|\,g)\geq \ln 2$'.
In order to show this, we will need the following Lemma:

\begin{Lemma}
For arbitrary objects $(p,g)$ and $(\tl{p},\tl{g})$
\[
(p,g)\geq (\tl{p},\tl{g})
\]
implies
\[
S(p\,\|\,g)\geq S(\tl{p}\,\|\,\tl{g})
\]
and
\[
S(g\,\|\,p)\geq S(\tl{g}\,\|\,\tl{p}).
\]
\end{Lemma}

\begin{Proof}
It is well-known that $S(.\|.)$ is a distance measure on the set of probability measures
which is decreasing with respect to stochastic maps (Uhlmann's monotonicity theorem \cite{OP}).
\end{Proof}

\begin{Corollary}(`Landauer's principle')
To obtain the perfectly initialized bit $\tl{p}=(0,1)$ from the maximally
unknown bit $\tl{g}=(1/2,1/2)$ one needs resources
$(p,g)$ with $S(p\,\|\,g)\geq \ln 2$, i.e,
\[
(p,g)\geq (\tl{p},\tl{g})
\]
implies
\[
S(p\,\|\,g)\geq \ln 2.
\]
\end{Corollary}

This can be seen by checking the equality $S(\tl{p}\,\|\,\tl{g})=\ln 2$.

\begin{Corollary}(`Perfect erasure is impossible with generic resources')
Let $p$ be  a state with $p_i\neq 0$ for every $i$.
Then there is no $n$ such that
\[
(p^{\otimes n},g^{\otimes n})\geq (\tl{p},\tl{g}),
\]
with $\tl{p}=(0,1)$ and $\tl{g}=(1/2,1/2))$.
This can be seen by $S(\tl{g}\,\|\,\tl{p})=\infty\neq
 S(g^{\otimes n}\,\|\,p^{\otimes n})$.
\end{Corollary}

Note that Landauer's principle is arguing with the relative entropy $S(p\,\|\,g)$
 whereas our analysis uses $S(g\,\|\,p)$.
This exchange of the role of $p$ and $g$ 
 is more important
than it seems: Both quantities measure the distance from the equilibrium state,
but with respect to the first distance measure the states 
$p:=(\epsilon,1-\epsilon)$
and $p':=(1,0)$ have almost the same distance from equilibrium if
$\epsilon$ is small. In contrast, the distance measure obtained by exchanging
the role between $p$ and $g$, converges to infinity  as $\epsilon$
tends to $0$.

 Therefore, $S(g\,\|\,p)$ seems more appropriate for describing
the difficulties in approaching the absolute Zero!
In other words: The usual thermodynamic quantities like
energy, free energy and entropy cannot explain `the hardness of the struggle
against the last milli-Kelvin above the absolute Zero'.

\section{What is the Kullback-Leibler information of a
 typical energy source?}

One may rephrase our results by the statement
`for reliable bit erasure one needs much more than the free
energy $\ln 2 kT$'. But this formulation is misleading:
Even for arbitrarily reliable bit erasure, there cannot exist
any lower bound tighter than the one given by  Landauer: If the  resource's
state is  a pure one it may enable perfect erasure even with the free
energy $\ln 2 kT$. This statement is trivial since any qubit can be
prepared into a perfect pure state if the resource is given by a qubit
with the identical energy gap as the first one 
being in a pure state. 
This example seems to be unserious since it shifts the problem of cooling
to the problem of supplying resources with the same temperature.
Furthermore the problem of cooling seems to have a circular logical structure.
Nevertheless the example shows, that any statements about tighter bounds
have to refer to particular assumptions about the statistical
properties of the energy source's state.

In view of this, we rephrase our results more carefully:
Given any resource object $O:=(\rho,\gamma)$ with the property that
no eigenvalue of $\rho$ is 0. Then an arbitrarily reliable bit erasure
process requires arbitrary many copies  of $O$, i.e., we need the resources
$O^n$ with appropriately large $n$, even if the free energy
of $O^m$ exceeds $\ln kT$ already for a considerably smaller number  $m$. 
For making definite statements about $n$ and $m$ one should make
assumptions about $\rho$ and $\gamma$ and fix the demanded error probability.
Deriving statistical properties  of the states of realistic energy sources
is not easy and should be a subject of further research.
However, from a quite fundamental point of view, it is quite
natural to ask for the `thermodynamic worth' of a {\em heat source}
with respect to good cooling and reliable bit erasure:
We assume that the resource's state $\rho$ is a thermal equilibrium state
with temperature $\tl{T}>T$.
This assumption is an example of a non-circular way of 
treating the problem of the required resources:
The resource's state is prepared by controlling {\em macroscopic} quantities
(in our example the temperature) without any direct possibility
of controlling its {\em microphysical} state.
We show that in our example the relative information can be
calculated explicitly  if the partition function
of the energy source is known:

\begin{Lemma}
Let $(p,g)$ be an object where $p$ is an equilibrium state 
for the inverse temperature $\tilde{\beta}$ and $g$ is the
equilibrium state for the environment's inverse temperature $\beta$.
With the partition function
\[
Z(\beta):=\sum_i e^{-\beta E_i}
\]
we have:
\[
p_i=e^{-\tl{\beta} E_i}/Z(\tl{\beta}),\,\, \, g_i=e^{-\beta}/Z(\beta).
\]
Hence
we get the Kullback-Leibler Information
\begin{eqnarray*}
S(g\,\|\,p)&=&\sum g_i \ln (g_i/p_i)\\&=&\ln Z(\tl{\beta}) -\ln Z(\beta) 
+ E_g(g) (\tl{\beta}-\beta),
\end{eqnarray*}
where $E_g(g)$ is the mean energy of the equilibrium state $g$.
This term is clearly finite for $\beta$ and $\tl{\beta}$ being finite.
Hence the required number of copies of the heat source
$(p,g)$ for cooling down to the demanded temperature can be estimated
 by knowing the temperatures and the partition function.
\end{Lemma}

\section{Conclusions}

To investigate the problem of cooling from a quite fundamental
point of view our model includes the driving energy source as a
quantum system with density matrix describing its statistical state.
This setup elucidated the lacks of the  traditional thermodynamic laws
for explaining the resource requirements for cooling processes
approaching the ground states:
It is by no means sufficient that the energy source
is able to supply  enough  free energy,
it rather is necessary, that the density matrix of the energy source
has a large enough distance from its equilibrium state
in another information theoretic
 sense. One has to distinguish between two different questions:
Firstly one wants to determine whether a qubit can be cooled down
even further if it is already colder than the environment's temperature.
This problem turned out to be essentially a geometric one and
we have shown, that the limit temperature at which every cooling process
breaks down is given by a simple parameter which we called
`maximal diagonal deviation' from equilibrium.
If one starts with a qubit having environment temperature,
 this limit temperature can in general only be approached by 
repeating a cooling procedure with refreshed resources
at each cycle.

The second problem is to determine the  temperature
which can be obtained by starting with a qubit with  
the environment's temperature if no such refreshment of
the resources is allowed.
Here the determination of the 
possibilities of cooling
 is essentially equivalent to the
 determination of an optimal decision rule
which can distinguish between the resources density matrix and
the corresponding equilibrium density matrix.
This result strongly emphasizes the fact that 
information theoretical arguments can rule out
physical processes in a way which goes far beyond
usual entropy arguments.

In a straightforward way, our theory applies to the 
more general question of the 
resources needed for preparing approximately
pure states in any multi-level quantum system.
This justifies the quite general formulation of the title:
The thermodynamic costs of {\em reliability}.

\section{Appendix}

For the proof of Lemma \ref{Bed} we need the following technical lemma:

\begin{Lemma}\label{Kuehnlein}
Let $A:=diag (a_1,\dots, a_n)$ with $a_1\geq a_2\geq \dots \geq a_n$
and $B:=diag (b_1,\dots,b_n)$
with $b_i=-1$ for $i\leq l$ and $b_i=1$ for $i>l$.

Let $u$ be an arbitrary unitary operator. Then 
\[
tr (A uBu^*)\geq tr (AB).
\]
\end{Lemma}

\begin{Proof}
We have:
\begin{eqnarray}\label{AuB}
&&tr (AuBu^*)-tr (AB)\nonumber \\
&=&\sum_j a_j \sum_i b_i (|u_{ji}|^2- \delta_{ij})\nonumber \\&=&
\sum_j a_j ( \sum_{i\neq j} b_i |u_{ji}|^2 -b_j \sum_{i\neq j} |u_{ji}|^2),
\end{eqnarray}
where we have used $\sum_j |u_{ji}|^2=1$ since $u$ is unitary. 
The term in equation (\ref{AuB}) reads as:
\begin{eqnarray}
\sum_j a_j \sum_{i\neq j} |u_{ji}|^2 (b_i-b_j)&=&\\
\sum_{j=1}^l a_j \sum_{i=l+1}^n |u_{ji}|^2 2+
\sum_{j=l+1}^n a_j \sum_{i=1}^l |u_{ji}|^2 (-2) &\geq&\\
2 a_l \sum_{j=1}^l \sum_{i=l+1}^n |u_{ji}|^2 -
2 a_{l+1} \sum_{j=l+1}^n\sum_{i=1}^l |u_{ji}|^2.
\end{eqnarray}
This term is greater or equal than zero since the double sums are the same:
Because $u$ is a unitary operator, the row square sums as well as the 
column  sums equal 1.
Therefore, 
\[
\sum_{j+1}^l \sum_{i=l+1}^n |u_{ji}|^2=l-\sum_{i,j \leq l} |u_{ji}|^2=
\sum_{j=l+1}^n \sum_{i=1}^l |u_{ji}|^2. 
\]
\end{Proof}

Now we are able to prove Lemma \ref{Bed}:

\begin{Proof} 
$u$ commutes with $\gamma \otimes \sigma$ and hence with its
spectral projections.
Therefore we have
\[
u=\sum u_j \,\, \hbox{ with }\,\, u_j:=\sum P_j u P_j. 
\]
Then it is sufficient to show
\[
tr (u_j \alpha u_j^* (1\otimes \sigma_z))\leq tr (P_j \alpha P_j 
(1\otimes \sigma_z)  ).
\] 
Since 
\[
tr (u_j \alpha^* u_j (1\otimes \sigma_z))=tr (u_j \alpha u^*_j P_j (1\otimes 
\sigma_z) P_j),
\]
we can reduce the problem
 completely to the situation of Lemma
\ref{Kuehnlein} by considering the  range of every  $P_j$ separately:
Restricted to  the range of $P_j$, the operator $u_j$ acts as a unitary one. 
\end{Proof}

\end{multicols}

\section*{Acknoledgements}

Thanks to 
S. K\"uhnlein for  the proof
of Lemma \ref{Kuehnlein} and to D. Lazic and R. Schack
for useful discussions.


\begin{thebibliography}{XXXX}
\bibitem{AAKVC} A. Aspect, E. Arimondo, R. Kaiser,
       N. Vansteenkiste, and C. Cohen-Tannoudji, Laser cooling
      below the one-photon recoil energy by
      velocity-selective coherent population trapping, 
      Phys. Rev. Lett. 61, No. 7, 826 (1988)


\bibitem{MECZ} G. Morigi, J. Eschner, J.I. Cirac, and P. Zoller,  Laser cooling of two trapped ions: sideband cooling beyond
                 the Lamb-Dicke limit, e-print
               quant-ph/9812014



\bibitem{MCLZ} G. Morigi, J.I. Cirac, M. Lewenstein, and P. Zoller,
              Ground state laser cooling beyond the Lamb-Dicke limit,
               e-print quant-ph/9706017



\bibitem{Be} C. Bennett, The thermodynamics of computation -- a review,
 Int. J. Theor. Phys., Vol. 21, (12), 1982   

\bibitem{La1} R. Landauer, Irreversibility and heat generation in the
          computing process,  IBM Res. J., July 1961.

\bibitem{La2} R. Landauer, Information is physical, Phys. Today, May 1991. 



\bibitem{AC} C. Adami, N. Cerf, Prolegomena to a non-equilibrium quantum
            statistical mechanics, e-print quant-ph/9904006



\bibitem{Pl} M. Plenio, The Holevo bound and Landauer's principle,
e-print quant-ph/9910086


\bibitem{BBPS} C. Bennett, H. Bernstein, S. Popescu, B. Schumacher,
 Concentrating partial entanglement by local operations, quant-ph/9511030


\bibitem{Ve} V. Vedral, Landauer's erasure, error correction
             and entanglement, e-print quant-ph/9903049   

\bibitem{BDSW} C. Bennett, D. DiVincenzo, J. Smolin, and
               W. Wootters, Mixed-state entanglement and quantum error
               correction, Phys. Rev. A, Vol. 54, No 5, Nov. 1996 

\bibitem{Ho} M. Horodecki, Limits for compression of
 quantum information carried
by ensemble of mixed states, Phys. Rev. A, Vol. 57, No. 5, May 1998.



\bibitem{SV} L. Schulman, U. Vazirani, Scalable NMR quantum computation,
    April 1998, e-print quant-ph/9804060 

\bibitem{Go} B. Gal-Or (editor), Modern developments in thermodynamics,
     John Wiley \& Sons, Israel University Press, Jerusalem 1974


\bibitem{Kr} U. Krengel, Ergodic Theory, Walter de Gruyter, Berlin 1985


\bibitem{CT} Th. Cover and J. Thomas, Elements of information theory,
             John Wiley \& Sons, Inc., New York 1991

             

\bibitem{Fu} C. Fuchs, Distinguishability and accessible information in 
              quantum theory,   Ph. D. thesis, University of New Mexico,
              December 1995, e-print quant-ph/9601020


\bibitem{Bl} R. Blahut, Principles and practice of information theory,
             Addison-Wesley Publishing Company, Reading, Massachusetts  1987


\bibitem{OP} M. Ohya and D. Petz, Quantum entropy and its use, Springer Verlag
             Berlin Heidelberg 1993

\end{thebibliography}
\end{document}